%% file: No-Regret_Forecasting_arXiv.tex
\RequirePackage[l2tabu, orthodox]{nag}
\documentclass[11pt,fleqn]{article}

\usepackage{footmisc}
\DefineFNsymbols{mySymbols}{{\ensuremath\dagger}}
\setfnsymbol{mySymbols}

\usepackage{setspace}
\input{Preamblesetup}
\usepackage{caption}
\usepackage{subfig}

\usepackage{algorithm}
\usepackage{algorithmic}

\title{No-Regret Forecasting with Egalitarian Committees}
\author{Jiun-Hua Su\thanks{
I thank Le-Yu Chen, Shiu-Sheng Chen, Yu-Chin Hsu, Chu-An Liu, and Shou-Yung Yin for helpful discussions.
Address correspondence to Jiun-Hua Su, 128 Academia Road, Section 2, Nankang, Taipei, 115 Taiwan; E-mail address: jhsu@econ.sinica.edu.tw.}}
\affil{Institute of Economics\\Academia Sinica}
\begin{document}
\maketitle
\thispagestyle{empty}

\begin{abstract}
The forecast combination puzzle is often found in literature:
The equal-weight scheme tends to outperform sophisticated methods of combining individual forecasts.
Exploiting this finding, we propose a hedge egalitarian committees algorithm (HECA),
which can be implemented via mixed integer quadratic programming.
Specifically, egalitarian committees are formed by the ridge regression with shrinkage toward equal weights;
subsequently, the forecasts provided by these committees are averaged by the hedge algorithm.
We establish the no-regret property of HECA.
Using data collected from the ECB Survey of Professional Forecasters,
we find the superiority of HECA relative to the equal-weight scheme during the COVID-19 recession.

\bigskip
\noindent
\textit{Keywords}: Hedge Egalitarian Committees Algorithm, No-Regret Forecasting, Mixed Integer Quadratic Programming, Forecast Combination Puzzle, ECB Survey of Professional Forecasters

\medskip
\noindent
\textit{JEL Classification}: C22, C52, C53

\end{abstract}

\fontsize{12}{18pt}\selectfont
\newpage
\setcounter{page}{1}

\section{Introduction}\label{Introduction}
Big data has been a buzzword in social science in recent years, and its popularity is witnessed in surveys such as \citet{Varian2014} and \citet{EinavLevin2014a} in economics, \citet{LazerRadford2017} in sociology, and \citet{Brady2019} in political science.
The importance of data in empirical studies is self-evident.
However, as argued by \citet{Lovell1983} almost forty years ago, ``it is by no means obvious that reductions in the costs of data mining have been matched by a proportional increase in our knowledge of how the economy actually works.''
The advance in theory is also important and indispensable for scientific progress.
Data science --- thus named perhaps because it emphasizes dialogues between theory and data --- suggests an effective way to improve knowledge.

A dialogue between theory and data has been exemplified by the forecast combination puzzle in econometrics.
This puzzle refers to the phenomenon that the equal-weight scheme, which is theoretically suboptimal in general, often outperforms the forecast combination with \citeauthor{BatesGranger1969}'s (\citeyear{BatesGranger1969}) optimal weights as well as other sophisticated combination methods in empirical studies.
The early dialogue had sparkled thought-provoking works on the combination of forecasts, as documented in \citet{Clemen1989}.
More recent surveys on the forecast combinations are provided by \citet{Timmermann2006} as well as \citet{ElliottTimmermann2016a}.
Indeed, the dialogue on such a puzzle continues to this day.
For example, acknowledging the equal-weight scheme as a high benchmark, \citet{DieboldShin2019} propose the \emph{egalitarian ridge regression}, which is the ridge regression with shrinkage toward the equal-weight scheme.\footnote{
Alternatively, \citet{DieboldPauly1990} propose empirical Bayes forecasting procedures with shrinkage toward the equal-weight scheme.
This Bayesian approach has been applied in, for example, \citet{StockWatson2004} and \citet{AiolfiTimmermann2006}.
}
Their idea has heuristic appeal but leaves open the question of an appropriate computational method.

In this paper, we fortify the theoretical and computational foundations of \citeauthor{DieboldShin2019}'s (\citeyear{DieboldShin2019}) shrinkage approach by developing a real-time forecasting algorithm under the decision-making framework given that the growing literature on forward-looking models in economics highlights the role of forecasting in decision making.\footnote{
As indicated in \citet{ClaridaGaliEtAl2000} and \citet{Mavroeidis2010}, a forecast-based interest rate rule in response to future macroeconomic conditions can provide a guideline for a monetary policy maker.
Forecasting matters in not only a public policy but a private agent's decision as well.
\citet{TanakaBloomEtAl2020} build a simple model concerning a firm's decisions on inputs under uncertainty to rationalize the empirical evidence that its GDP forecast accuracy is a predictor of profitability and productivity.
}
A decision maker first organizes \emph{egalitarian committees}, that is, committees providing their forecasts, respectively, via a ridge regression with shrinkage toward the simple average of individual forecasts selected by mixed integer quadratic programming (MIQP).
The application of MIQP and partition of a parameter space solve \citeauthor{DieboldShin2019}'s computational difficulty in egalitarian ridge regression with simultaneous selection of individual forecasts.
Next, the decision maker pools committee forecasts by applying a variant of \citeauthor{FreundSchapire1997}'s (\citeyear{FreundSchapire1997}) hedge algorithm.
This variant depends on an estimate of the maximal committee loss for the duration of its implementation.
The decision maker's two-stage implementation of real-time forecasting is referred to as \emph{hedge egalitarian committees algorithm}, henceforth abbreviated to HECA.

We establish non-asymptotic upper bounds on the \emph{average regret} attained by HECA, which is the decision maker's average forecasting loss in excess of the smallest average forecasting loss accomplished by these egalitarian committees.
First, these upper bounds indicate the decision maker's own acumen of business cycles could pay off because given the committee forecasts, a more precise estimate of the maximal committee loss \textit{ceteris paribus} yields a tighter upper bound on the average regret.
Furthermore, these upper bounds show that HECA has \emph{no-regret} property; that is, the decision maker's long-run performance should be at least as good as the best long-run performance accomplished by the egalitarian committees.
This result is in line with the findings in the online learning literature.
An excellent overview of this literature is recently provided by \citet{Cesa-BianchiOrabona2021}.
More importantly, this no-regret property implies the superiority of HECA relative to the equal-weight scheme in the long run.
It is such `theoretical' superiority that makes HECA eligible for the competition with the equal-weight scheme; however, its `empirical' superiority remains to be examined.

To examine whether HECA outperforms the equal-weight scheme in an empirical study, we focus on the quarterly one-year-ahead forecasts of Euro-area real GDP growth in Survey of Professional Forecasters (SPF), which is conducted by the European Central Bank (ECB).
The purpose of selecting this dataset is twofold.
On the one hand, it generates the equal-weight scheme that has particularly hard-to-beat forecasting performance, as demonstrated in \citet{GenreKennyEtAl2013}, \citet{ConflittiDeMolEtAl2015}, and \citet{DieboldShin2019}.
On the other hand, it involves not-so-big data such that theoretical parts of data science (inclusive of domain knowledge, statistical methods, and computational techniques) are crucially important.
Our empirical results show that HECA keeps pace with the equal-weight scheme before the outbreak of COVID-19 but wins the competition during the COVID-19 recession.
Despite the superiority of HECA relative to the equal-weight scheme, HECA suffers from an upsurge in forecasting loss around the onset of COVID-19 pandemic.
This pattern is consistent with previous research, as indicated in \citet{ChauvetPotter2013}.
We also find that the formation of egalitarian committees gives HECA an advantage over \citeauthor{FreundSchapire1997}'s hedge algorithm during the COVID-19 recession.

In addition to the pursuit of forecasting performance, we are dedicated to credible forecasting in a spirit that only credible assumptions are maintained, as emphasized in \citet{Manski2013a}.
To achieve this goal, we treat the data generating process (DGP) of target variables and their individual forecasts as a \emph{black box}; that is, no assumption on such DGP is imposed.
Instead, the proposed HECA is a data-driven and adaptive approach: At each round, it outputs a combined forecast with more weights on forecasts provided by committees that have performed well in the past; furthermore, the built-in updating mechanism enables HECA to adapt to the environment in the presence of structural breaks that may make forecasters' relative performance unstable over time.
The unstable performance is called \emph{model instability} in literature, and an excellent survey of this issue is provided by \citet{Rossi2013}.
Additionally, the committee forecasts, as inputs of HECA, are obtained by the rolling egalitarian ridge regression scheme.
The rolling scheme is used to guard against possible parameter drift, as indicated by \citet{West2006}, whereas the shrinkage toward the simple average of selected individual forecasts is supported by empirical evidence in literature.

HECA embodies interdisciplinary research, which is another marked characteristic of data science.\footnote{
This interdisciplinary characteristic is vividly illustrated in Drew Conway's data science Venn diagram, which can be found at \url{http://drewconway.com/zia/2013/3/26/the-data-science-venn-diagram}.
}
Knowledge in econometric literature, statistical methods, numerical and computational techniques, and online learning modeling are woven into HECA for the decision maker's real-time forecasting.
The empirical findings in econometrics treat the equal-weight scheme as a high benchmark, to which the weights should shrink.
In spite of alleviating numerical instability, the ridge regression in general fails to achieve the simultaneous selection of individual forecasts, which is implemented by mixed integer optimization.
The hedge algorithm further allows for the adaptability to sequential data in real time.
Equipped with these designs, HECA complements, but does not replace, existing forecasting methods.
It is particularly useful in the situation where the decision maker has limited access to predictors.
For surveys of data-rich methods, the reader is referred to \citet{StockWatson2006} and \citet{ChauvetPotter2013}.

Throughout this paper, we write $\Norm{z}_{1}$, $\Norm{z}_{2}$, and $\Norm{z}_{\infty}$ for the one-norm, two-norm, and infinity-norm of a generic column vector $z$ in an Euclidean space, respectively.
We denote the collection of positive integers and the collection of real numbers by $\mathbb{N}$ and $\mathbb{R}$, respectively.

The structure of the remaining paper is as follows.
Section~\ref{Data} describes features of data collected from the ECB SPF and Eurostat.
Section~\ref{Methodology} presents the organization of egalitarian committees, the decision maker's HECA, and the theoretical upper bounds on the average regret of HECA.
Section~\ref{EmpiricalResults} discusses the empirical results of applying HECA to real-time forecasting of the year-on-year growth rate of euro area.
Section~\ref{conclusion} concludes.
Technical proofs are deferred to the appendix.

\section{Data from Eurostat and ECB SPF}\label{Data}
The forecast target variables in this paper are the year-on-year Euro-area GDP growth estimates collected from Eurostat, the European Statistical Agency.\footnote{
These estimates are available at
\url{https://ec.europa.eu/eurostat/web/national-accounts/data/other}.
}
Due to data revisions, several estimates for a given quarter are released by Eurostat.
Following \citet{GenreKennyEtAl2013}, we focus on the $t+45$ flash estimates, which are published about 45 days after the associated quarter, for our empirical study in Section~\ref{EmpiricalResults}.
The evaluation sample runs from the first quarter of 2012 to the third quarter of 2020.

As in \citet{GenreKennyEtAl2013}, \citet{ConflittiDeMolEtAl2015}, and \citet{DieboldShin2019}, we focus on the quarterly one-year-ahead forecasts of Euro-area real GDP growth in the ECB SPF.
These one-year-ahead forecasts, however, are actually six to eight months ahead.
For example, in the questionnaire for the third quarter of 2018,
macroeconomic experts participating in the SPF are asked for the expected year-on-year real GDP growth for the first quarter of 2019 and provided with the GDP growth for the first quarter of 2018 as a reference.

One noticeable feature in the SPF is the frequent entry, exit, and reentry of experts so that an unbalanced panel arises.
As pointed out in \citet{GenreKennyEtAl2013}, such an unbalanced panel may yield sampling distortions.
To lessen the extent of undesirable distortions, we exclude experts who did not reply in two consecutive quarters during the evaluation period spanning from the first quarter of 2012 to the third quarter of 2020.
After this removal, there remain $21$ experts.
Hereafter, we focus on the forecasts provided by this filtered panel of experts.
In Figure~\ref{Figure1_new}, we mark a slot by the notation x if a forecast is provided by an associated expert for a specific quarter in the SPF; otherwise, we leave it blank.
We further replace each missing value of a forecast with the simple average of the rest of the reported forecasts for the same quarter.
For example, Expert 038 provides forecasts throughout the evaluation period except the one for the third quarter of 2015; this missing forecast is filled in with the simple average of forecasts provided by the other 19 experts, as the forecast associated with Expert 110 is also unreported.

Another well-known feature in the SPF is the forecast combination puzzle:
It is hard for other sophisticated schemes to improve on the performance of equal-weight scheme, as indicated by \citet{GenreKennyEtAl2013} and \citet{ConflittiDeMolEtAl2015}.
Some rationales behind this puzzle are proposed in literature.
From a theoretical perspective, as shown in \citet{Timmermann2006}, the equal-weight scheme is optimal if individual forecast errors have the same variance and identical pairwise correlation.
From a practical perspective, as noted in \citet{SmithWallis2009} and \citet{ConflittiDeMolEtAl2015}, the finite-sample error and numerical instability may make the estimated optimal weights inferior to the equal-weight scheme in terms of forecasting performance.

Both perspectives are crucial in an empirical study using the evaluation sample from the SPF.
First, Table~\ref{Table1_new} and Figure~\ref{Figure2_new} show the sample variances and pairwise correlation coefficients, respectively, of individual forecast errors for the $21$ experts in the filtered panel.
Since the sample variances are similar to each other whereas the sample correlation coefficients are centered around $0.995$, the hard-to-beat performance of equal-weight scheme is unsurprising.
In addition, the asymptotic approximation of estimated weights is arguably imprecise because the overall sample size --- $35$ quarters --- is obviously small.
More importantly, the numerical instability in the estimated optimal weights is severe; for example, the condition number associated with the ordinary least square regression using the evaluation sample is $31,112$.\footnote{
The (2-norm) condition number of a matrix $A$ is defined as
\begin{align*}
\kappa_{2}(A)
\equiv\frac{\max\limits_{b:\Norm{b}_{2}=1}\Norm{Ab}_{2}}
{\min\limits_{b:\Norm{b}_{2}=1}\Norm{Ab}_{2}}
\end{align*}
and often used to evaluate the stability of a linear system in numerical analysis.
As argued in \citet{BelsleyKuhEtAl1980}, ``moderate to strong relations are associated with condition indexes of $30$ to $100$.''}
This finding yields a clue as to the application of ridge regression, a classical approach in literature to alleviating numerical instability, to the estimation of optimal weights.
Recognizing the remarkable performance of equal-weight scheme,
\citet{DieboldShin2019} propose the egalitarian ridge regression, which is the ridge regression with shrinkage toward the equal-weight scheme.
Following their approach, we further develop an algorithm in the next section that can select experts in each quarter and achieve some satisfactory objective in hindsight.

\section{Forecasting with Egalitarian Committees}\label{Methodology}
The fundamental importance of economic forecasting for forward-looking private agents and public policy makers motivates us to propose an algorithm that incorporates features of the SPF forecasts and outperforms the equal-weight scheme under the decision-making framework.
Roughly speaking, we consider the situation where a single decision maker is allowed access to forecasts provided by anonymous experts using either quantitative models or model-free judgments, and such forecasts generating processes are unknown to the decision maker.\footnote{
These experts' potentially strategic behaviors are also ignored by the decision maker.
For the strategic forecasting, we refer the reader to \citet{MarinovicOttavianiEtAl2013} and references cited therein.
}
This decision maker is assumed to minimize the cumulative squared loss without discounting.
The squared loss can be replaced with other loss functions, for example those documented in Section 2.2 of \citet{ElliottTimmermann2016a}, in the rest of this paper.
We refrain from this replacement because the squared loss is used in common with the literature on forecast combination puzzle.

To elaborate on the proposed method, we now introduce notation.
Suppose that there are $M$ experts providing a forecast of the target variable $y_{t}$, respectively, before its realization.
These individual forecasts are denoted by $f_{t}\equiv(f_{t,1},\dots,f_{t,M})^{\top}$, where $f_{t, m}$ stands for the forecast of $y_{t}$ provided by expert $m\in\{1,\dots,M\}$.\footnote{
In our empirical analysis in Section~\ref{EmpiricalResults}, the vector $f_{t}$ of forecasts in the SPF are six to eight months prior to the release of the $t + 45$ flash estimate $y_{t}$ by Eurostat.
}
Accessing the data encompassing current forecasts, realized target variables, and their corresponding forecasts, the decision maker announces his or her own forecast of $y_{t}$ by a two-stage method:
At the first stage, the decision maker imagines $M$ committees $\{\mathcal{C}_{c}\}_{c=1}^{M}$, where committee $\mathcal{C}_{c}$ consists of $c$ members selected among $M$ experts;
subsequently, each committee provides a forecast $\hat{y}_{t,c}$, which is a combination of forecasts provided by its $c$ members.
At the second stage, this decision maker applies the hedge algorithm to $\{\hat{y}_{t,c}\}_{c=1}^{M}$ and then yields his or her own forecast of $y_{t}$.
To complete the two-stage method, we explain how these committees $\{\mathcal{C}_{c}\}_{c=1}^{M}$ are formed at the first stage and how the hedge algorithm works at the second stage in the subsections below.

\subsection{Egalitarian Committees}\label{EgalitarianCommittee}
The imaginary committee $\mathcal{C}_{c}$ is organized by solving the following optimization problem for a fixed rolling window $r\in \mathbb{N}$ and every $\lambda$ in a pre-specified set $\Lambda$ of grids:\footnote{
As indicated in \citeauthor{ElliottTimmermann2016a} (\citeyear{ElliottTimmermann2016a}, p.\ 378), the length of estimation window can be selected by the cross-validation method, which is however rarely done.
}
\begin{enumerate}
\item[(P1)]
\begin{align*}
&\min_{b\in\mathbb{R}^{M}}
\sum_{s=l}^{l+r-1}\left[y_{t-s}-f^{\top}_{t-s}b\right]^{2}
+\lambda\Norm*{b-\frac{1}{\Norm{b}_{0}}\bm{1}}^{2}_{2}\\[0.2cm]
&\hspace{0.2cm}\text{s.t.}\;
0\leq b_{j}\leq 1,\;\; \text{for}\; j=1,\dots,M;\\[0.1cm]
&\hspace{0.9cm}\sum_{j=1}^{M}b_{j}=1;\\[0.1cm]
&\hspace{0.9cm}\Norm{b}_{0}=c\in\mathbb{N},
\end{align*}
\end{enumerate}
where $\Norm{b}_{0}$ is the number of nonzero elements in $b\equiv(b_{1},\dots,b_{M})^{\top}$ and $\bm{1}$ is the $M$ dimensional column vector of ones.
The rolling scheme is adopted to guard against possible parameter drift.
Since all individual forecasts are measured on the same scale, they are not standardized in this ridge-type regression.
In addition, the lag term can be set to be either $l=1$ or $l=2$ for real-time forecasting with the SPF forecasts.
Let $\hat{\beta}_{t,c}(\lambda)$ be a solution to problem (P1) associated with $\lambda$, and $\iota_{m}\in \mathbb{R}^{M}$ be a unit vector with $m$-th element equal to one.
The tuning parameter $\hat{\lambda}_{t,c}$ is selected by setting
\begin{align*}
\hat{\lambda}_{t,c}
&\equiv\arg\min_{\lambda\in\Lambda}\sum_{s=l}^{l+r_{\lambda}-1}
\left[y_{t-s}-f_{t-s}^{\top}\hat{\beta}_{t-s,c}(\lambda)\right]^{2},
\end{align*}
where $r_{\lambda}\in \mathbb{N}$ denotes the number of periods for validation.
The set
\begin{align*}
\mathcal{C}_{c}\equiv \left\{m: \iota^{\top}_{m}\hat{\beta}_{t,c}(\hat{\lambda}_{t,c})>0\right\}
\end{align*}
is called the \emph{egalitarian committee} with $c$ members, for problem (P1) can be viewed as a subproblem of partial egalitarian ridge regression in \citet{DieboldShin2019}.
To see this, let
\begin{align*}
\tilde{\beta}_{t}(\lambda)
\equiv\arg\min
\left\{\sum_{s=l}^{l+r-1}\left[y_{t-s}-f^{\top}_{t-s}b\right]^{2}
+\lambda\Norm*{b-\frac{1}{\Norm{b}_{0}}\bm{1}}^{2}_{2}:
b\in\left\{\hat{\beta}_{t,c}(\lambda)\right\}_{c=1}^{M}\right\},
\end{align*}
where $\hat{\beta}_{t,c}(\lambda)$ is a minimizer of (P1) for each $c\in\{1,\dots,M\}$ and a fixed $\lambda$.
Although the objective function is discontinuous due to $\Norm{b}_{0}$, we have
\begin{align*}
\tilde{\beta}_{t}(\lambda)=\arg&\min_{b\in\mathbb{R}^{M}}
\sum_{s=l}^{l+r-1}\left[y_{t-s}-f^{\top}_{t-s}b\right]^{2}
+\lambda\Norm*{b-\frac{1}{\Norm{b}_{0}}\bm{1}}^{2}_{2}\\[0.2cm]
&\text{s.t.}\;
0\leq b_{j}\leq 1,\;\; \text{for}\; j=1,\dots,M;\\[0.1cm]
&\hspace{0.5cm}\sum_{j=1}^{M}b_{j}=1.
\end{align*}
Phrased differently, the partition of a parameter space according to the value of $\Norm{b}_{0}$ allows us to recover $\tilde{\beta}_{t}(\lambda)$.
Therefore, \citeauthor{DieboldShin2019}'s (\citeyear{DieboldShin2019}) `one-step' partial egalitarian ridge regression can be equivalently implemented as long as problem (P1) is successfully solved for every $c$.\footnote{
Although \citeauthor{DieboldShin2019}'s (\citeyear{DieboldShin2019}) partial egalitarian ridge regression concerns the inclusion of \citeauthor{Tibshirani1996}'s (\citeyear{Tibshirani1996}) one-norm regularization in the objective function rather than in the constraints, the idea of partitioning a parameter space still works \textit{mutatis mutandis}.
}

To solve problem (P1), we recast it as the following MIQP:
\begin{enumerate}
\item[(P2)]
\begin{align*}
&\min_{b\in\mathbb{R}^{M},\; d\in\mathbb{R}^{M}}
\sum_{s=l}^{l+r-1}\left[y_{t-s}-f^{\top}_{t-s}b\right]^{2}
+\lambda\Norm*{b-\frac{1}{c}\bm{1}}^{2}_{2}\\[0.2cm]
&\hspace{0.5cm}\text{s.t.}\;
d_{j}\epsilon\leq b_{j}\leq d_{j},\;\; \text{for}\; j=1,\dots,M;\\[0.2cm]
&\hspace{1.15cm}d_{j}\in\{0,1\},\;\;\hspace{0.43cm} \text{for}\; j=1,\dots,M;\\[0.2cm]
&\hspace{1cm}\sum_{j=1}^{M}b_{j}=1;\;\;\sum_{j=1}^{M}d_{j}=c\in\mathbb{N}.
\end{align*}
\end{enumerate}
If $\epsilon$ is the smallest machine-representable positive real number,\footnote{
As defined in \citeauthor{Judd1998} (\citeyear{Judd1998}, p.\ 30), a machine zero is referred to as a quantity equivalent to zero on a machine.
The positive real number $\epsilon$ is not a machine zero, but every positive real number less than $\epsilon$ is a machine zero.
}
then problems (P1) and (P2) are computationally equivalent.
The intuition is that under the constraints $d_{j}\epsilon\leq b_{j}\leq d_{j}$ and $d_{j}\in\{0,1\}$, the dummy variable $d_{j}=\Ind{[b_{j}>0]}$ indicates whether $b_{j}$ is positive; that is, expert $j$ is selected in the committee $\mathcal{C}_{c}$ with size $\Norm{b}_{0}$, which is equal to the sum of $d_{j}$'s.
We summarize the discussion in the following proposition.
\begin{Pro}\label{Pro1}
Suppose that $\epsilon$ is the smallest machine-representable positive real number.
The optimization problem (P1) and the MIQP (P2) are computationally equivalent in the following sense:
\begin{enumerate}[(i)]
\item If the machine yields the minimizer $b^{*}$ of (P1), then $(b^{*}, d^{*})$ is a minimizer of (P2), where $d^{*}\equiv(d_{1}^{*}, \dots, d_{M}^{*})^{\top}$ and $d_{j}^{*}=\Ind{[b^{*}_{j}>0]}$ for each $j=1,\dots, M$.
\item If the machine yields the minimizer $(b^{*}, d^{*})$ of (P2), then $b^{*}$ is a minimizer of (P1).
\end{enumerate}
\end{Pro}

\vspace{0.3cm}
A conceptually simple method of solving problem (P2) is exhaustive enumeration.
To see this, note that there are $\binom{M}{c}$ feasible choices of $d\equiv(d_{1},\dots,d_{M})^{\top}$ in problem (P2).
For any given feasible $d$, this optimization problem is essentially the constrained ridge regression with the unknown parameter $b$.
Implementing these $\binom{M}{c}$ ridge regressions thus suffices to solve problem (P2).
We call this approach \emph{complete subset ridge regressions} by analogy with complete subset regressions in \citet{ElliottGarganoEtAl2013}.
This exhaustive approach, however, may be computationally inefficient because \emph{every} egalitarian committee is asked to provide a forecast $\hat{y}_{t,c}\equiv f_{t}^{\top}\hat{\beta}_{t,c}(\hat{\lambda}_{t,c})$; consequently, there are $\binom{M}{1}+\binom{M}{2}+\dots+\binom{M}{M}=2^{M}-1$ ridge regressions to be carried out for every $\lambda\in\Lambda$.

Instead of such an exhaustive search for $\{\hat{\beta}_{t,c}(\lambda)\}_{c=1}^{M}$ in the parameter space, the modern solver Gurobi can be used to implement the MIQP in problem (P2).
The practical tractability of moderate-size MIQP, though NP-hard in nature, can be attributed to the rapid advances in computation power.
According to \citet{BertsimasDunn2019}, the overall speedup, inclusive of solvers and hardware, is approximately two trillion between 1991 and 2016.
The advance of mixed integer optimization has sparked recent studies in statistics and econometrics, for example \citet{BertsimasKingEtAl2016} and \citet{ChenLee2018}, among others.
Details about algorithmic developments of mixed integer optimization can be found in \citet{JuengerLieblingEtAl2010}, \citet{ConfortiCornuejolsEtAl2014}, and references cited therein.

\begin{Rmk}
The constraint on total weight ($b^{\top}\bm{1}=1$) and the range constraints ($0\leq b_{j}\leq 1$ for each $j$) are included in problems (P1) and (P2) because they have the following nice properties.
First, as pointed out in \citet{GrangerRamanathan1984}, the combined forecast $\hat{y}_{t,c}\equiv f_{t}^{\top}\hat{\beta}_{t,c}(\hat{\lambda}_{t,c})$ provided by the egalitarian committee $\mathcal{C}_{c}$, under the constraint on total weight, remains unbiased if every member in this committee provides an unbiased forecast of $y_{t}$.
As shown in \citet{Diebold1988}, the constraint on total weight also implies that the forecast errors $y_{t}-\hat{y}_{t,c}$ are serially uncorrelated if the individual forecast errors made by the members in this committee are serially uncorrelated.
In addition, \citet{JagannathanMa2003} indicate the shrinkage effect of range constraints on reducing the estimation error of the covariances of experts' forecast errors.
Finally, \citet{ConflittiDeMolEtAl2015} suggest that these constraints together can improve the numerical stability in computation.
\end{Rmk}

In econometrics and statistics, assumptions about the DGP of $(y_{t}, f_{t}^{\top})$ are usually imposed to establish theoretically nice properties of a forecasting method.
\citeauthor{ElliottTimmermann2016a} (\citeyear{ElliottTimmermann2016a}, p.\ 320), however, put it this way:
\begin{quote}
\emph{Interestingly, combination methods that attempt to explicitly model time variation in the combination weights often fail to perform well, suggesting that regime switching or model ``breakdown'' can be difficult to predict or even track through time.}
\end{quote}
Additionally, the knowledge about how individual forecasts are generated by experts is unknown in principle to the decision maker.
A practical example given in \citet{Diebold2015} is that forecasts are purchased from a vendor using proprietary models, which are not revealed to the decision maker.
Recognizing such limits to knowledge, the decision maker attempts to neither model nor assume the DGP of $(y_{t}, f_{t}^{\top})$ and is dedicated to credible forecasting.\footnote{
As argued in \citet{Manski2013a}, ``[t]he fundamental difficulty of empirical research is to decide what assumptions to maintain.''
He further argues that ``[s]tronger assumptions yield conclusions that are more powerful but less credible.''
}
This decision maker thus deals with the real-time forecasting problem by the hedge algorithm based on the past performance of the egalitarian committees, as described in the next subsection.

\subsection{Hedge Egalitarian Committees Algorithm}\label{Hedge}
After receiving forecasts $\hat{y}_{t}\equiv(\hat{y}_{t,1},\dots,\hat{y}_{t,M})^{\top}$ made by all  egalitarian committees, the decision maker announces his or her own forecast, which is a weighted average of $\{\hat{y}_{t,c}\}_{c=1}^{M}$.
Subsequently, the nature announces the realization of the target variable $y_{t}$.
The sequence of target variables can be generated as in statistical models.
For example, it can represent business cycles undulating along a trend, either deterministic or stochastic;
it can exhibit structural breaks with changing points, either known or unknown;
and it can describe switching among different states, either observed or unobservable.
Further examples about statistical modeling can be found in \citet{Pesaran2015} and \citet{PenaTsay2021}.
Alternatively, this sequence of target variables can be adversarially generated as in game-theoretic models, where the nature attempts to maximize the decision maker's forecasting loss.
Details about game-theoretic analysis can be found in \citet{Cesa-BianchiLugosi2006} and \citet{SchapireFreund2012}.
Briefly, the following happen in order for each round $t$:
\begin{enumerate}
\item[1.] Egalitarian committees announce their forecast combinations
\begin{align*}
\hat{y}_{t,c}\equiv f_{t}^{\top}\hat{\beta}_{t,c}(\hat{\lambda}_{t,c}),
\;\text{for}\;
c=1,\dots,M,
\end{align*}
where $\hat{\beta}_{t,c}$ and $\hat{\lambda}_{t,c}$ are obtained by the method in the previous subsection;
\item[2.] Decision maker announces the forecast
\begin{align*}
\hat{\hat{y}}_{t}\equiv \pi_{t}^{\top}\hat{y}_{t}
\end{align*}
according to some distribution $\pi_{t}\in\triangle^{M}$;
\item[3.] Nature announces $y_{t}\in\mathbb{R}$;
\item[4.] Egalitarian committees incur loss
\begin{align*}
\ell_{t,c}\equiv \left(y_{t}-\hat{y}_{t,c}\right)^{2},
\;\;\text{for}\;
c=1,\dots,M.
\end{align*}
\end{enumerate}
Adopting the aforementioned strategy (i.e., announcement of $\hat{\hat{y}}_{t}$ in each round) the decision maker \emph{ex ante} aims to obtain small average regret
\begin{align*}
R_{T}\equiv \frac{1}{T}\sum_{t=1}^{T}\left(y_{t}-\hat{\hat{y}}_{t}\right)^{2}
-\min_{c\in\{1,\dots,M\}}\frac{1}{T}\sum_{t=1}^{T}\left(y_{t}-\hat{y}_{t,c}\right)^{2},
\end{align*}
by selecting the sequence $\{\pi_{t}\}_{t=1}^{T}$ of distributions.
To achieve this goal, this decision maker selects $\{\pi_{t}\}_{t=1}^{T}$ by HECA, whose pseudocode is shown as Algorithm~\ref{HECA}.

\begin{algorithm}[t]
\setstretch{2}
\caption{Hedge Egalitarian Committees Algorithm}
\label{HECA}
\begin{algorithmic}[1]
\REQUIRE ~~\\
    $\pi_{1}=\pi_{2}\equiv(1/M,\dots,1/M)^{\top}\in \triangle^{M}$;\\
    $\omega_{1,c}=\omega_{2,c}\equiv 1$ for each $c=1,\dots,M$;\\
    $\eta_{1}=\frac{2}{B_{1}}\sqrt{\log\{M\}}$, where $B_{1}$ is an assumed maximal committee loss;\\
\ENSURE ~~\\
	A sequence $\{\pi_{t}\}_{t=1}^{T}$ of distributions;\\
\FOR{each $t=3,\dots, T$}
    \STATE collect $\ell_{(t-2),c}$ for each $c=1,\dots,M$;\\
    \STATE $\omega_{t,c}\leftarrow \omega_{(t-2),c}
    \exp\{-\eta_{(t-2)}\ell_{(t-2),c}\}\;$ for each $c=1,\dots,M$;\\
	\STATE $\pi_{t,c}\leftarrow \frac{\omega_{t,c}}{\sum_{m=1}^{M}\omega_{t,m}}\;$
    for each $c=1,\dots, M$;\\
    \STATE $B_{(t-1)}\leftarrow \max\left\{B_{(t-2)}, \max_{c=1,\dots,M}\ell_{(t-2),c}\right\}$;\\
    \STATE $\eta_{(t-1)}\leftarrow \frac{2}{B_{(t-1)}}\sqrt{\frac{\log\{M\}}{t-1}}$;\\
\ENDFOR
\end{algorithmic}
\end{algorithm}

HECA, adapted from the hedge algorithm in \citet{FreundSchapire1997}, incorporates features of the decision maker's real-time forecasting based on the SPF forecasts.
Suppose that the decision maker announces his or her forecast $\hat{\hat{y}}_{t}$ immediately after receiving the SPF forecasts.
In this case, the decision maker does not know the realized forecasting loss $\{\ell_{t,c}\}_{c=1}^{M}$ until round $t+2$.
Hence, in the first two rounds, the decision maker has no information about any committee's performance and thus uses the uniform distributions $\pi_{1}$ and $\pi_{2}$.
In the third and ensuing rounds, the decision maker observes every committee's performance $\{\ell_{(t-2),c}\}_{c=1}^{M}$ of two rounds prior, thereby updating the distribution $\pi_{t}\equiv(\pi_{t,1},\dots,\pi_{t,M})^{\top}$.
Note that HECA requires an estimate of the maximal committee loss throughout the $T$ rounds.
The decision maker assumes $B_{1}$ to be this maximal loss, which might be a biased estimate, in the first two rounds, and updates it subsequently round by round.

HECA reflects the presumption embraced by the decision maker: A committee with relatively better performance (i.e., smaller $\ell_{(t-2)}$) would maintain the momentum to perform relatively well in the current round; therefore, its weight $\pi_{t,c}$ in the combined forecast $\hat{\hat{y}}_{t}$ should relatively increase.
The idea of performance-based pooling of forecasts has been used in econometrics, for example forecasts weighted by inverse mean squared error in \citet{StockWatson1999} and \citet{CapistranTimmermann2009}, and aggregated forecast through exponential reweighting in \citet{Yang2004} and \citet{WeiYang2012}, among others.
The built-in updating mechanism makes HECA adaptive to the environment.
The adaptability of HECA is in sharp contrast to the constancy of equal-weight scheme even in the possibly ever-changing environment.

The following theorem gives upper bounds on the decision maker's average regret.
These upper bounds are non-asymptotic; that is, they hold for every finite $T\in\mathbb{N}$.
\begin{Thm}\label{Thm1}
Let $\bar{B}_{T}\equiv\max\{\ell_{t,c}: t=1,\dots,T;\; c=1,\dots,M\}$.
HECA guarantees that for all $M,T\in\mathbb{N}$,
\begin{align*}
R_{T}\leq
\begin{cases}
(1+2\gamma_{\text{u}})\bar{B}_{T}\sqrt{\frac{\log\{M\}}{T}}, & \text{if $\bar{B}_{T}=\gamma_{\text{u}} B_{1}$ for some real number $\gamma_{\text{u}}> 1$};\\[0.3cm]
3\gamma_{\text{o}}\bar{B}_{T}\sqrt{\frac{\log\{M\}}{T}}, & \text{if $B_{1}=\gamma_{\text{o}} \bar{B}_{T}$ for some real number $\gamma_{\text{o}}> 1$};\\[0.3cm]
3\bar{B}_{T}\sqrt{\frac{\log\{M\}}{T}},& \text{otherwise}.
\end{cases}
\end{align*}
\end{Thm}
\vspace{0.3cm}
\noindent
The non-asymptotic upper bounds in Theorem~\ref{Thm1} hold \emph{without} requirement of any assumption on the DGP.
The decision maker may underestimate or overestimate the maximal loss $\bar{B}_{T}$.
The biased estimation, however, is a contributing factor of an upper bound on the average regret.
A sharper upper bound can be obtained if the magnitude of underestimation (overestimation), measured by $\gamma_{\text{u}}$ ($\gamma_{\text{o}}$), is smaller.
It turns out that given the same committee forecasts and duration of implementing HECA, decision makers having different evaluations of business cycle may achieve different forecasting performances even if they use HECA.
Calculation of these upper bounds is also easy \textit{ex post}.
In contrast, upper bounds in \citet{Yang2004} and \citet{WeiYang2012} involve nuisance parameters of the underlying DGP and need further evaluation.

Moreover, if the sequence $\{\bar{B}_{T}\}_{T=1}^{\infty}$ is bounded above,\footnote{
The monotonicity of $\{\bar{B}_{T}\}_{T=1}^{\infty}$, together with its boundedness, implies that $\lim_{T\to\infty}\bar{B}_{T}<\infty$.
}
then HECA exhibits no regret because these upper bounds on the average regret all shrink to zero as $T$ tends to infinity.
Phrased differently, the performance of HECA is at least close to that of the best egalitarian committee in the long run.
The no-regret property is common in the online learning literature.
We refer the reader to \citeauthor{Cesa-BianchiLugosi2006}'s (\citeyear{Cesa-BianchiLugosi2006}) monograph for early findings and \citeauthor{Cesa-BianchiOrabona2021}'s (\citeyear{Cesa-BianchiOrabona2021}) survey for recent advances.
From a pragmatic standpoint, the assumption about boundedness of $\{\bar{B}_{T}\}_{T=1}^{\infty}$ could be inconsequential.
To see this, note that
\begin{align*}
\bar{B}_{T}
\leq \max\left\{(y_{t}-f_{t,i})^{2}: t=1,\dots,T;\; i=1,\dots,M\right\}
\end{align*}
by Jensen's inequality; additionally, as indicated by \citeauthor{ElliottTimmermann2016a} (\citeyear{ElliottTimmermann2016a}, p.\ 17), ``[i]n practice, forecasts are usually bounded and extremely large forecasts typically get trimmed as they are deemed implausible.''

More importantly, Theorem~\ref{Thm1} establishes the intuition that the decision maker's two-stage method should outperform the equal-weight scheme whenever $T$ is so large that the upper bounds are small.
Theorem~\ref{Thm1} implies that in the long run, HECA should perform at least as well as the best egalitarian committee.
In addition, the best egalitarian committee would dominate the egalitarian committee $\mathcal{C}_{M}$, which should in turn be weakly better than the equal-weight scheme.
It follows from these arguments that HECA would outweigh the equal-weight scheme in terms of long-run forecasting performance.

If the decision maker regularly postpones announcing $\hat{\hat{y}}_{t}$ until the realization of $y_{(t-1)}$,
then the information on $\{\ell_{(t-1),c}\}_{c=1}^{M}$ could be exploited.
Because of the updated information, the decision maker's forecasting performance is expected to improve.
Indeed, HECA with delayed announcements (Algorithm~\ref{HECA_delayed_announcements}) allows for tighter upper bounds on the average regret, as shown in the following theorem.

\begin{Thm}\label{Thm2}
HECA with delayed announcements guarantees that for all $M,T\in\mathbb{N}$,
\begin{align*}
R_{T}\leq
\begin{cases}
\frac{(1+2\gamma_{\text{u}})}{\sqrt{2}}\bar{B}_{T}\sqrt{\frac{\log\{M\}}{T}}, & \text{if $\bar{B}_{T}=\gamma_{\text{u}} B_{1}$ for some real number $\gamma_{\text{u}}> 1$};\\[0.3cm]
\frac{3\gamma_{\text{o}}}{\sqrt{2}} \bar{B}_{T}\sqrt{\frac{\log\{M\}}{T}}, & \text{if $B_{1}=\gamma_{\text{o}} \bar{B}_{T}$ for some real number $\gamma_{\text{o}}> 1$};\\[0.3cm]
\frac{3}{\sqrt{2}}\bar{B}_{T}\sqrt{\frac{\log\{M\}}{T}},& \text{otherwise},
\end{cases}
\end{align*}
where $\bar{B}_{T}$ is defined in Theorem~\ref{Thm1}.
\end{Thm}

\begin{algorithm}[t]
\setstretch{2}
\caption{Hedge Egalitarian Committees Algorithm with Delayed Announcements}
\label{HECA_delayed_announcements}
\begin{algorithmic}[1]
\REQUIRE ~~\\
    $\pi_{1}\equiv(1/M,\dots,1/M)^{\top}\in \triangle^{M}$;\\
    $\omega_{1,c}\equiv 1$ for each $c=1,\dots,M$;\\
    $\eta_{1}=\frac{1}{B_{1}}\sqrt{2\log\{M\}}$, where $B_{1}$ is an assumed maximal committee loss;\\
\ENSURE ~~\\
	A sequence $\{\pi_{t}\}_{t=1}^{T}$ of distributions;\\
\FOR{each $t=2,\dots, T$}
    \STATE collect $\ell_{(t-1),c}$ for each $c=1,\dots,M$;\\
    \STATE $\omega_{t,c}\leftarrow \omega_{(t-1),c}
    \exp\{-\eta_{(t-1)}\ell_{(t-1),c}\}\;$ for each $c=1,\dots,M$;\\
	\STATE $\pi_{t,c}\leftarrow \frac{\omega_{t,c}}{\sum_{m=1}^{M}\omega_{t,m}}\;$
    for each $c=1,\dots, M$;\\
    \STATE $B_{t}\leftarrow \max\left\{B_{t-1}, \max_{c=1,\dots,M}\ell_{(t-1),c}\right\}$;\\
    \STATE $\eta_{t}\leftarrow \frac{1}{B_{t}}\sqrt{\frac{2\log\{M\}}{t}}$;\\
\ENDFOR
\end{algorithmic}
\end{algorithm}
\vspace{0.3cm}
\noindent

\begin{Rmk}
Unlike HECA, which updates $w_{t,c}$ according to the latest observed loss $\ell_{(t-2), c}$, an alternative updating mechanism makes smooth the adjustment to new weights by setting
\begin{align}\label{EFP-ECA}
\omega_{t,c}\leftarrow \omega_{(t-2),c}
\exp\left\{-\frac{\eta_{(t-2)}}{t-2}\sum_{\tau=1}^{t-2}\ell_{\tau,c}\right\}\; \text{for each $c=1,\dots,M$}.
\end{align}
This updating mechanism relies on the latest observed empirical loss $(t-2)^{-1}\sum_{\tau=1}^{t-2}\ell_{\tau,c}$ and parallels \citeauthor{FudenbergLevine1995}'s (\citeyear{FudenbergLevine1995}) \emph{exponential fictitious play}, which has the no-regret property, also known as \emph{Hannan-consistency} in the literature on learning in games.
For a book length treatment of this topic, see \citet{FudenbergLevine1998} and \citet{Cesa-BianchiLugosi2006}.
Similarly, we can consider the exponential fictitious play with delayed announcements by replacing (\ref{EFP-ECA}) with the following updating mechanism
\begin{align}\label{EFP-ECA2}
\omega_{t,c}\leftarrow \omega_{(t-1),c}
\exp\left\{-\frac{\eta_{(t-1)}}{t-1}\sum_{\tau=1}^{t-1}\ell_{\tau,c}\right\}\; \text{for each $c=1,\dots,M$}.
\end{align}
\end{Rmk}

\section{An Empirical Study on the Growth Rate of Euro Area}\label{EmpiricalResults}
Convinced of the theoretically asymptotic performance of HECA, we are now concerned with its forecasting performance for the evaluation sample mentioned in Section~\ref{Data}.

We first concentrate on the competition among the equal-weight scheme, HECA (Algorithm~\ref{HECA} with $l=2$) and HECA with delayed announcements (Algorithm~\ref{HECA_delayed_announcements} with $l=1$).
The latter two algorithms involve the first-stage MIQP, which is implemented by the Gurobi Python interface, and the tuning parameters are set by $r=16$, $r_{\lambda}=1$, $\Lambda=\{0.01g\}_{g=1}^{200}$, and $\epsilon=5\times 10^{(-324)}$.\footnote{The number $5\times 10^{(-324)}$ is equal to the product of \emph{sys.float\_info.min} and \emph{sys.float\_info.epsilon} in Python 3.7, and the mixed integer optimization are carried out by Gurobi 9.0.3, which is available at \url{https://www.gurobi.com/}.
}
We set $B_{1}$ to be the maximum of individual forecasting losses (observed by the decision maker) from the first quarter of 2012 to the quarter prior to $t=1$.
HECA, in comparison to its counterpart with delayed announcements, requires two extra rounds for `in-sample' estimation of $\hat{\beta}$'s and validation of $\hat{\lambda}$'s.
Thus, we consider $t=1$ in HECA to be the fourth quarter of 2016 and $t=1$ in HECA with delayed announcements to be the second quarter of 2016.

Table~\ref{Table2_new} reports their forecasting losses and associated differences per round for this competition.
As can be seen, HECA and the equal-weight scheme are nearly neck and neck until the first quarter of 2020.
Keeping abreast of the equal-weight scheme, which is a well-known high benchmark, HECA also performs well.
HECA further outperforms the equal-weight scheme since the second quarter of 2021, in which ``the fall in economic activity was unprecedented in depth, speed and scope''.\footnote{
This description is given in the news published on 29th March 2021 by Euro Area Business Cycle Dating Committee. Further details are available at
\url{https://eabcn.org/sites/default/files/eabcdc_findings_29_march_2021.pdf}.
}
HECA with delayed announcements even achieves better forecasting performance by exploiting updated information.
Given the small-sample survey data from the SPF, we do not pursue statistical testing for the superiority of HECA because existing tests accounting for in-sample estimation error, for example the tests developed in \citet{DieboldMariano1995} and \citet{GiacominiWhite2006}, rely on out-of-sample asymptotic approximation to determine the critical value.

Moreover, Figure~\ref{Figure3_new} implies that the number of experts in the committee performing best in a single round is not constant but time-varying.
Despite this instability, the theoretical results in Section~\ref{Methodology} suggest that HECA could perform as well as the best committee over the entire evaluation period in hindsight.
As can be seen from Table~\ref{Table3_new}, the average regret is relatively small given the substantial impact of COVID-19 pandemic on the euro area economy.
It is worth noting that the best egalitarian committee in the fourth quarter of 2019 and in the third quarter of 2020 are identical, and the average regret is less than $0.03$ if HECA with/without delayed announcements terminates in the fourth quarter of 2019.

Finally, we turn the spotlight on the cousin and ancestor of HECA.
As shown in Table~\ref{Table4_new}, the exponential fictitious play with updating mechanism~(\ref{EFP-ECA}) performs almost the same as HECA; similarly, the exponential fictitious play with updating mechanism~(\ref{EFP-ECA2}) very much resembles HECA with delayed announcements in terms of forecasting ability.
The close resemblance gives a hint on the no-regret property of exponential fictitious play.
Table~\ref{Table5_new} shows that although neither \citeauthor{FreundSchapire1997}'s (\citeyear{FreundSchapire1997}) hedge algorithm nor HECA dominates each other before the fourth quarter of 2019,
HECA wins the competition since the first quarter of 2020.
Thus, combining the results in Tables~\ref{Table2_new} and~\ref{Table5_new}, we find that during the COVID-19 recession, the formation of egalitarian committees gives HECA a competitive edge over the hedge algorithm, which beats the equal-weight scheme by adaptability.

\section{Conclusion}\label{conclusion}
The proposed HECA should be in the data scientist's toolkit for three reasons as follows.
From a theoretical perspective, HECA outputs credible forecasting because it relies on practically convincing assumptions and meanwhile achieves an asymptotically negligible upper bound on the average regret.
From an empirical perspective, HECA outweighs the equal-weight scheme after the outbreak of COVID-19 in euro area, whereas the equal-weight scheme only outperforms HECA by a margin, if any, before such an outbreak.
From a methodological perspective, HECA differs from other data-rich methods in that it is applicable in the context where no extra predictor of the target variables, except for the forecasts provided by the experts, is available for the decision maker.

We do not deal with the optimal timing of implementing HECA.
Our empirical results seem to suggest that compared with the equal-weight scheme, HECA would be suitable for forecasting around business cycle turning points.
It is also unclear whether the duration of implementing HECA should be determined at the very beginning.
We delegate these fascinating issues for future work.

\begin{appendices}
\numberwithin{equation}{section}

\section{Technical proofs}\label{AppendixA}

\subsection{Proof of Proposition~\ref{Pro1}}
For ease of notation, we define functions
$Q_{1}:\mathbb{R}^{M}\to\mathbb{R}$ and $Q_{2}:\mathbb{R}^{M}\to\mathbb{R}$ to be
\begin{align*}
Q_{1}(b)
\equiv\Norm{Y-Fb}_{2}^{2}+\lambda\Norm*{b-\frac{1}{\Norm{b}_{0}}\bm{1}}_{2}^{2}\;\;\text{and}\;\;
Q_{2}(b)
\equiv\Norm{Y-Fb}_{2}^{2}+\lambda\Norm*{b-\frac{1}{c}\bm{1}}_{2}^{2}.
\end{align*}
It is clear that $Q_{1}(b)=Q_{2}(b)$ whenever $\Norm{b}_{0}=c$.
\begin{enumerate}[(i)]
\item First, we show that $(b^{*},d^{*})$ is feasible for problem (P2).
Since $b^{*}$ satisfies the constraints in (P1), $0\leq b_{j}^{*}\leq 1$ for every $j$.
By the construction of $d_{j}^{*}$,
\begin{align*}
\sum_{j=1}^{M}d_{j}^{*}
=\sum_{j=1}^{M}\Ind{[b_{j}^{*}>0]}
=\Norm{b^{*}}_{0}
=c
\end{align*}
and $d_{j}^{*}\in\{0,1\}$ for each $j$.
It remains to establish $d_{j}^{*}\epsilon\leq b_{j}^{*}\leq d_{j}^{*}$ for each $j$.
Note that $b_{j}^{*}$ is either zero or positive.
If the machine yields $b_{j}^{*}=0$, then $d_{j}^{*}=0$ and
\begin{align*}
d_{j}^{*}\epsilon
=0
\leq b_{j}^{*}
\leq 0
=d_{j}^{*}.
\end{align*}
If the machine yields $b_{j}^{*}>0$, then
\begin{align*}
d_{j}^{*}\epsilon
=\epsilon
\leq b_{j}^{*}
\leq 1
=d_{j}^{*}
\end{align*}
because $\epsilon$ is the smallest machine-representable positive real number.

Next, we prove that $(b^{*},d^{*})$ is a minimizer of problem (P2) by contradiction.
Suppose that $(\tilde{b},\tilde{d})$ is feasible for problem (P2) and $Q_{2}(\tilde{b})<Q_{2}(b^{*})$.
It follows from the constraints in (P2) that
\begin{align*}
\Norm{\tilde{b}}_{0}
=\sum_{j=1}^{M}\Ind{[\tilde{b}_{j}>0]}
=\sum_{j=1}^{M}\tilde{d}_{j}
=c.
\end{align*}
Thus, $\tilde{b}$ is feasible for problem (P1).
It turns out that
\begin{align*}
Q_{1}(\tilde{b})
=Q_{2}(\tilde{b})
<Q_{2}(b^{*})
=Q_{1}(b^{*}),
\end{align*}
contradicting the assumption that $b^{*}$ is a minimizer of problem (P1).
\item First, we show that $b^{*}$ is feasible for problem (P1).
Since $(b^{*},d^{*})$ satisfies the constraints in (P2), we have
\begin{align*}
0
\leq d_{j}^{*}\epsilon
\leq b_{j}^{*}
\leq d_{j}^{*}
\leq 1
\end{align*}
for each $j$ and
\begin{align*}
\Norm{b^{*}}_{0}
=\sum_{j=1}^{M}\Ind{[b_{j}^{*}>0]}
=\sum_{j=1}^{M}d_{j}^{*}=c.
\end{align*}
Thus, $b^{*}$ satisfies the constraints in (P1).

Next, we prove that $b^{*}$ is a minimizer of (P1) by contradiction.
Suppose that $\check{b}$ is computationally feasible for problem (P1) and $Q_{1}(\check{b})<Q_{1}(b^{*})$.
Let $\check{d}_{j}\equiv\Ind{[\check{b}_{j}>0]}$ for each $j$.
By construction,
\begin{align*}
\check{d}_{j}\in\{0,1\}
\;\;\text{and}\;\;
\check{d}_{j}\epsilon\leq \check{b}_{j}\leq \check{d}_{j}
\end{align*}
for each $j$.
Additionally, we have
\begin{align*}
\sum_{j=1}^{M}\check{d}_{j}
=\sum_{j=1}^{M}\Ind{[\check{b}_{j}>0]}
=\Norm{\check{b}}_{0}
=c.
\end{align*}
Hence, $(\check{b},\check{d})$ is computationally feasible for problem (P2).
It follows that
\begin{align*}
Q_{2}(\check{b})
=Q_{1}(\check{b})
<Q_{1}(b^{*})
=Q_{2}(b^{*}),
\end{align*}
contradicting the assumption that $(b^{*},d^{*})$ is a minimizer of (P2).
\end{enumerate}

\subsection{Proof of Theorem~\ref{Thm1}}
Without loss of generality, let $\eta_{t}=\frac{2}{B_{t}}\sqrt{\frac{\log\{M\}}{t}}$ and $B_{t+1}=\max\left\{B_{t}, \max_{c=1,\dots,M}\ell_{t,c}\right\}$ for every $t\in\mathbb{N}$.
We denote the Kullback-Leibler divergence of $x=(x_{1},\dots,x_{M})^{\top}$ and $z=(z_{1},\dots,z_{M})^{\top}$ by
\begin{align*}
D(x\|z)=\sum_{c=1}^{M}x_{c}\log\left\{\frac{x_{c}}{z_{c}}\right\} \;\;\text{for any}\;\;
(x,z)\in\triangle^{M}\times \text{ri}(\triangle^{M}),
\end{align*}
where $\text{ri}(\triangle^{M})$ is the relative interior of probability $M$-simplex $\triangle^{M}\subseteq \mathbb{R}^{M}$.
For each $t\in\mathbb{N}$, HECA outputs
\begin{align*}
\pi_{t+2}=\arg\min_{\pi\in\triangle^{M}}
\left[\langle\ell_{t}, \pi-\pi_{t}\rangle+\frac{1}{\eta_{t}}D(\pi\|\pi_{t})\right].
\end{align*}

The Bregman proximal inequality given in Lemma 3.1 of \citet{Teboulle2018} implies that for any $\pi\in\triangle^{M}$ and $t\in\mathbb{N}$,
\begin{align*}
\langle\ell_{t}, \pi_{t}-\pi\rangle
\leq\frac{1}{\eta_{t}}\left[D(\pi\|\pi_{t})-D(\pi\|\pi_{t+2})\right]
+\langle\ell_{t}, \pi_{t}-\pi_{t+2}\rangle-\frac{1}{\eta_{t}}D(\pi_{t+2}\|\pi_{t}).
\end{align*}
In addition, by Pinsker's inequality,
we have
\begin{align*}
D(\pi\|\pi_{t})\geq \frac{1}{2}\Norm{\pi-\pi_{t}}_{1}^{2},
\end{align*}
for all $\pi\in\triangle^{M}$.
It follows from the two inequalities above that for any $\pi\in\triangle^{M}$ and $t\in\mathbb{N}$,
\begin{align*}
\langle\ell_{t}, \pi_{t}-\pi\rangle
&\leq\frac{1}{\eta_{t}}\left[D(\pi\|\pi_{t})-D(\pi\|\pi_{t+2})\right]
+\Norm{\ell_{t}}_{\infty}\Norm{\pi_{t+2}-\pi_{t}}_{1}
-\frac{1}{2\eta_{t}}\Norm{\pi_{t+2}-\pi_{t}}_{1}^{2}\\
&\leq\frac{1}{\eta_{t}}\left[D(\pi\|\pi_{t})-D(\pi\|\pi_{t+2})\right]
+\frac{\eta_{t}}{2}\Norm{\ell_{t}}_{\infty}^{2}.
\end{align*}
Since $\eta_{t}$ is decreasing in $t$ and $\Norm{\ell_{t}}_{\infty}\leq B_{t+1}$,
\begin{align*}
\sum_{t=1}^{T}\langle\ell_{t}, \pi_{t}\rangle-\sum_{t=1}^{T}\langle\ell_{t}, \pi\rangle
&\leq \frac{1}{\eta_{T}}\sum_{t=1}^{T}\left[D(\pi\|\pi_{t})-D(\pi\|\pi_{t+2})\right]
+ \frac{1}{2}\sum_{t=1}^{T}\eta_{t}\Norm{\ell_{t}}_{\infty}^{2}\\
&\leq \frac{1}{\eta_{T}}\left[D(\pi\|\pi_{1})+D(\pi\|\pi_{2})\right]
+\frac{1}{2}\sum_{t=1}^{T}\eta_{t}B_{t+1}^{2}\\
&\leq \frac{2\log\{M\}}{\eta_{T}}
+\frac{1}{2}\sum_{t=1}^{T}\eta_{t}B_{t+1}^{2},
\end{align*}
for any $T\in\mathbb{N}$.
Substituting $\eta_{t}=\frac{2}{B_{t}}\sqrt{\frac{\log\{M\}}{t}}$ into the last inequality yields
\begin{align}\label{InequalityA1}
\sum_{t=1}^{T}\langle\ell_{t}, \pi_{t}\rangle-\sum_{t=1}^{T}\langle\ell_{t}, \pi\rangle
&\leq B_{T}\sqrt{T\log\{M\}}
+\sqrt{\log\{M\}}\sum_{t=1}^{T}\frac{B_{t+1}^{2}}{B_{t}\sqrt{t}}.
\end{align}
We first establish the upper bound on $R_{T}$ for the case of $\bar{B}_{T}>B_{1}$.
The right hand side of Inequality~(\ref{InequalityA1}) is bounded above by
\begin{align}\label{InequalityA2}
\bar{B}_{T}\sqrt{T\log\{M\}}
+\gamma_{\text{u}} \bar{B}_{T}\sqrt{\log\{M\}}\int_{0}^{T}\frac{1}{\sqrt{s}}\Myd s
= (1+2\gamma_{\text{u}})\bar{B}_{T}\sqrt{T\log\{M\}}
\end{align}
because $B^{2}_{t+1}/B_{t}\leq \bar{B}^{2}_{T}/B_{1}=\gamma_{\text{u}}\bar{B}_{T}$ for each $t=1,\dots,T$.
By Jensen's inequality, we have
\begin{align}\label{InequalityA3}
\frac{1}{T}\sum_{t=1}^{T}\left(y_{t}-\hat{\hat{y}}_{t}\right)^{2}
&=\frac{1}{T}\sum_{t=1}^{T}\left(\sum_{c=1}^{M}\pi_{t,c}(y_{t}-\hat{y}_{t,c})\right)^{2}\notag\\
&\leq\frac{1}{T}\sum_{t=1}^{T}\sum_{c=1}^{M}\pi_{t,c}(y_{t}-\hat{y}_{t,c})^{2}\notag\\
&=\frac{1}{T}\sum_{t=1}^{T}\langle\ell_{t},\pi_{t}\rangle.
\end{align}
Combining Inequalities~(\ref{InequalityA1})-(\ref{InequalityA3}), we obtain
\begin{align*}
R_{T}\leq\frac{1}{T}\sum_{t=1}^{T}\langle\ell_{t}, \pi_{t}\rangle
-\min_{\pi\in\triangle^{M}}\frac{1}{T}\sum_{t=1}^{T}\langle\ell_{t}, \pi\rangle
\leq(1+2\gamma_{\text{u}})\bar{B}_{T}\sqrt{\frac{\log\{M\}}{T}}.
\end{align*}
Let us now move on to the remaining two cases.
In both cases, $B_{t}=B_{1}$ for all $t$.
It follows that the right hand side of Inequality~(\ref{InequalityA1}) is bounded above by
\begin{align}\label{InequalityA4}
B_{1}\sqrt{T\log\{M\}}+B_{1}\sqrt{\log\{M\}}\int_{0}^{T}\frac{1}{\sqrt{s}}\Myd s
&=3B_{1}\sqrt{T\log\{M\}}.
\end{align}
Combining Inequalities~(\ref{InequalityA1}), (\ref{InequalityA3}) and (\ref{InequalityA4}) yields
\begin{align*}
R_{T}
\leq 3B_{1}\sqrt{\frac{\log\{M\}}{T}}.
\end{align*}
We complete the proof by noting that $B_{1}=\gamma_{\text{o}}\bar{B}_{T}$ and $B_{1}=\bar{B}_{T}$ correspond to the second and third case in the statement of this theorem, respectively.

\subsection{Proof of Theorem~\ref{Thm2}}
A simple modification of the proof of Theorem~\ref{Thm1} yields the results.

\end{appendices}

\normalsize
\bibliography{No-Regret_Forecasting}
\bibliographystyle{ecta}

\include{Figure1_new}
\include{Figure2_new}
\include{Figure3_new}

\include{Table1_new}
\include{Table2_new}
\include{Table3_new}
\include{Table4_new}
\include{Table5_new}

\end{document}

%% file: Preamblesetup.tex
\usepackage{microtype}
\usepackage{mathtools,amssymb,amsthm,latexsym} 
\usepackage{mathrsfs} 
\usepackage{bbm,bm} 
\usepackage{natbib} 
\usepackage[shortlabels]{enumitem}
\usepackage{graphicx} 
\usepackage{booktabs} 
\usepackage{threeparttable} 
\usepackage[table]{xcolor} 
\usepackage[letterpaper,top=2.9cm,bottom=2.4cm,hmargin={3cm,3cm}]{geometry}
\usepackage{verbatim} 
\usepackage{tikz} 
   \usetikzlibrary{patterns,decorations.pathreplacing}
\usepackage{hyperref} 
   \hypersetup{colorlinks=true, linkcolor=blue, citecolor=blue!50!black, urlcolor=cyan}
\usepackage[capitalize,noabbrev]{cleveref}
\usepackage[titletoc,title]{appendix}
\usepackage{titling}
\usepackage{authblk} 
\usepackage{setspace}

\theoremstyle{definition}

\newtheorem*{Def*}{Definition}
\theoremstyle{plain}

\newtheorem*{Lem*}{Lemma}
\newtheorem{Pro}{Proposition}
\newtheorem{Thm}{Theorem}
\newtheorem*{Thm*}{Theorem}

\newtheorem{Rmk}{Remark}
\newtheorem*{Rmk*}{Remark}

\newcommand{\Myd}{\;\mathrm{d}} 
\newcommand{\Ind}[1]{\mathbbm{1}_{#1}} 

\DeclarePairedDelimiter\Norm{\lVert}{\rVert} 

%% file: Figure1_new.tex
\begin{figure}[t]
\caption{Entry and Exit of Experts}\label{Figure1_new}
\vspace{0.3cm}
\begin{tikzpicture}[xscale=0.37,yscale=0.35]
\draw [<->] (36,0) node[right]{\tiny{Quarter}} -- (0,0) -- (0,22)
node[left]{\tiny{Expert}};

\draw [thin] (-0.1,1) node[left]{\tiny{004}} -- (0.1,1);
\draw [thin] (-0.1,2) node[left]{\tiny{006}} -- (0.1,2);
\draw [thin] (-0.1,3) node[left]{\tiny{015}} -- (0.1,3);
\draw [thin] (-0.1,4) node[left]{\tiny{016}} -- (0.1,4);
\draw [thin] (-0.1,5) node[left]{\tiny{020}} -- (0.1,5);
\draw [thin] (-0.1,6) node[left]{\tiny{022}} -- (0.1,6);
\draw [thin] (-0.1,7) node[left]{\tiny{023}} -- (0.1,7);
\draw [thin] (-0.1,8) node[left]{\tiny{024}} -- (0.1,8);
\draw [thin] (-0.1,9) node[left]{\tiny{037}} -- (0.1,9);
\draw [thin] (-0.1,10) node[left]{\tiny{038}} -- (0.1,10);
\draw [thin] (-0.1,11) node[left]{\tiny{039}} -- (0.1,11);
\draw [thin] (-0.1,12) node[left]{\tiny{048}} -- (0.1,12);
\draw [thin] (-0.1,13) node[left]{\tiny{052}} -- (0.1,13);
\draw [thin] (-0.1,14) node[left]{\tiny{085}} -- (0.1,14);
\draw [thin] (-0.1,15) node[left]{\tiny{089}} -- (0.1,15);
\draw [thin] (-0.1,16) node[left]{\tiny{095}} -- (0.1,16);
\draw [thin] (-0.1,17) node[left]{\tiny{096}} -- (0.1,17);
\draw [thin] (-0.1,18) node[left]{\tiny{098}} -- (0.1,18);
\draw [thin] (-0.1,19) node[left]{\tiny{107}} -- (0.1,19);
\draw [thin] (-0.1,20) node[left]{\tiny{110}} -- (0.1,20);
\draw [thin] (-0.1,21) node[left]{\tiny{112}} -- (0.1,21);

\draw [ultra thick] (1,-0.2)
node[align=center, below]{\fontsize{4pt}{2cm}\selectfont{2012}\\[-0.25cm]\fontsize{4pt}{2cm}\selectfont{Q1}} -- (1,0.2);
\draw [thin] (2,-0.1) -- (2,0.1);
\draw [thin] (3,-0.1) -- (3,0.1);
\draw [thin] (4,-0.1) -- (4,0.1);
\draw [ultra thick] (5,-0.2)
node[align=center, below]{\fontsize{4pt}{2cm}\selectfont{2013}\\[-0.25cm]\fontsize{4pt}{2cm}\selectfont{Q1}} -- (5,0.2);
\draw [thin] (6,-0.1) -- (6,0.1);
\draw [thin] (7,-0.1) -- (7,0.1);
\draw [thin] (8,-0.1) -- (8,0.1);
\draw [ultra thick] (9,-0.2)
node[align=center, below]{\fontsize{4pt}{2cm}\selectfont{2014}\\[-0.25cm]\fontsize{4pt}{2cm}\selectfont{Q1}} -- (9,0.2);
\draw [thin] (10,-0.1) -- (10,0.1);
\draw [thin] (11,-0.1) -- (11,0.1);
\draw [thin] (12,-0.1) -- (12,0.1);
\draw [ultra thick] (13,-0.2)
node[align=center, below]{\fontsize{4pt}{2cm}\selectfont{2015}\\[-0.25cm]\fontsize{4pt}{2cm}\selectfont{Q1}} -- (13,0.2);
\draw [thin] (14,-0.1) -- (14,0.1);
\draw [thin] (15,-0.1) -- (15,0.1);
\draw [thin] (16,-0.1) -- (16,0.1);
\draw [ultra thick] (17,-0.2)
node[align=center, below]{\fontsize{4pt}{2cm}\selectfont{2016}\\[-0.25cm]\fontsize{4pt}{2cm}\selectfont{Q1}} -- (17,0.2);
\draw [thin] (18,-0.1) -- (18,0.1);
\draw [thin] (19,-0.1) -- (19,0.1);
\draw [thin] (20,-0.1) -- (20,0.1);
\draw [ultra thick] (21,-0.2)
node[align=center, below]{\fontsize{4pt}{2cm}\selectfont{2017}\\[-0.25cm]\fontsize{4pt}{2cm}\selectfont{Q1}} -- (21,0.2);
\draw [thin] (22,-0.1) -- (22,0.1);
\draw [thin] (23,-0.1) -- (23,0.1);
\draw [thin] (24,-0.1) -- (24,0.1);
\draw [ultra thick] (25,-0.2)
node[align=center, below]{\fontsize{4pt}{2cm}\selectfont{2018}\\[-0.25cm]\fontsize{4pt}{2cm}\selectfont{Q1}} -- (25,0.2);
\draw [thin] (26,-0.1) -- (26,0.1);
\draw [thin] (27,-0.1) -- (27,0.1);
\draw [thin] (28,-0.1) -- (28,0.1);
\draw [ultra thick] (29,-0.2)
node[align=center, below]{\fontsize{4pt}{2cm}\selectfont{2019}\\[-0.25cm]\fontsize{4pt}{2cm}\selectfont{Q1}} -- (29,0.2);
\draw [thin] (30,-0.1) -- (30,0.1);
\draw [thin] (31,-0.1) -- (31,0.1);
\draw [thin] (32,-0.1) -- (32,0.1);
\draw [ultra thick] (33,-0.2)
node[align=center, below]{\fontsize{4pt}{2cm}\selectfont{2020}\\[-0.25cm]\fontsize{4pt}{2cm}\selectfont{Q1}} -- (33,0.2);
\draw [thin] (34,-0.1) -- (34,0.1);
\draw [thin] (35,-0.1) -- (35,0.1);

\node (004,2012Q1) at (1,1) {x};
\node (006,2012Q1) at (1,2) {x};
\node (015,2012Q1) at (1,3) {x};
\node (016,2012Q1) at (1,4) {x};
\node (020,2012Q1) at (1,5) {x};
\node (023,2012Q1) at (1,7) {x};
\node (024,2012Q1) at (1,8) {x};
\node (037,2012Q1) at (1,9) {x};
\node (038,2012Q1) at (1,10) {x};
\node (048,2012Q1) at (1,12) {x};
\node (052,2012Q1) at (1,13) {x};
\node (085,2012Q1) at (1,14) {x};
\node (089,2012Q1) at (1,15) {x};
\node (095,2012Q1) at (1,16) {x};
\node (096,2012Q1) at (1,17) {x};
\node (098,2012Q1) at (1,18) {x};
\node (107,2012Q1) at (1,19) {x};
\node (110,2012Q1) at (1,20) {x};
\node (112,2012Q1) at (1,21) {x};

\node (004,2012Q2) at (2,1) {x};
\node (006,2012Q2) at (2,2) {x};
\node (015,2012Q2) at (2,3) {x};
\node (016,2012Q2) at (2,4) {x};
\node (020,2012Q2) at (2,5) {x};
\node (022,2012Q2) at (2,6) {x};
\node (023,2012Q2) at (2,7) {x};
\node (024,2012Q2) at (2,8) {x};
\node (037,2012Q2) at (2,9) {x};
\node (038,2012Q2) at (2,10) {x};
\node (039,2012Q2) at (2,11) {x};
\node (048,2012Q2) at (2,12) {x};
\node (052,2012Q2) at (2,13) {x};
\node (085,2012Q2) at (2,14) {x};
\node (089,2012Q2) at (2,15) {x};
\node (095,2012Q2) at (2,16) {x};
\node (096,2012Q2) at (2,17) {x};
\node (098,2012Q2) at (2,18) {x};
\node (107,2012Q2) at (2,19) {x};
\node (110,2012Q2) at (2,20) {x};
\node (112,2012Q2) at (2,21) {x};

\node (004,2012Q3) at (3,1) {x};
\node (006,2012Q3) at (3,2) {x};
\node (015,2012Q3) at (3,3) {x};
\node (016,2012Q3) at (3,4) {x};
\node (020,2012Q3) at (3,5) {x};
\node (022,2012Q3) at (3,6) {x};
\node (023,2012Q3) at (3,7) {x};
\node (024,2012Q3) at (3,8) {x};
\node (037,2012Q3) at (3,9) {x};
\node (038,2012Q3) at (3,10) {x};
\node (039,2012Q3) at (3,11) {x};
\node (048,2012Q3) at (3,12) {x};
\node (052,2012Q3) at (3,13) {x};
\node (085,2012Q3) at (3,14) {x};
\node (089,2012Q3) at (3,15) {x};
\node (095,2012Q3) at (3,16) {x};
\node (096,2012Q3) at (3,17) {x};
\node (098,2012Q3) at (3,18) {x};
\node (107,2012Q3) at (3,19) {x};
\node (110,2012Q3) at (3,20) {x};
\node (112,2012Q3) at (3,21) {x};

\node (004,2012Q4) at (4,1) {x};
\node (006,2012Q4) at (4,2) {x};
\node (015,2012Q4) at (4,3) {x};
\node (016,2012Q4) at (4,4) {x};
\node (020,2012Q4) at (4,5) {x};
\node (022,2012Q4) at (4,6) {x};
\node (023,2012Q4) at (4,7) {x};
\node (024,2012Q4) at (4,8) {x};
\node (037,2012Q4) at (4,9) {x};
\node (038,2012Q4) at (4,10) {x};
\node (039,2012Q4) at (4,11) {x};
\node (052,2012Q4) at (4,13) {x};
\node (085,2012Q4) at (4,14) {x};
\node (089,2012Q4) at (4,15) {x};
\node (095,2012Q4) at (4,16) {x};
\node (098,2012Q4) at (4,18) {x};
\node (107,2012Q4) at (4,19) {x};
\node (110,2012Q4) at (4,20) {x};
\node (112,2012Q4) at (4,21) {x};

\node (006,2013Q1) at (5,2) {x};
\node (015,2013Q1) at (5,3) {x};
\node (016,2013Q1) at (5,4) {x};
\node (020,2013Q1) at (5,5) {x};
\node (022,2013Q1) at (5,6) {x};
\node (023,2013Q1) at (5,7) {x};
\node (024,2013Q1) at (5,8) {x};
\node (037,2013Q1) at (5,9) {x};
\node (038,2013Q1) at (5,10) {x};
\node (039,2013Q1) at (5,11) {x};
\node (048,2013Q1) at (5,12) {x};
\node (052,2013Q1) at (5,13) {x};
\node (085,2013Q1) at (5,14) {x};
\node (089,2013Q1) at (5,15) {x};
\node (095,2013Q1) at (5,16) {x};
\node (096,2013Q1) at (5,17) {x};
\node (098,2013Q1) at (5,18) {x};
\node (107,2013Q1) at (5,19) {x};
\node (110,2013Q1) at (5,20) {x};
\node (112,2013Q1) at (5,21) {x};

\node (004,2013Q2) at (6,1) {x};
\node (006,2013Q2) at (6,2) {x};
\node (015,2013Q2) at (6,3) {x};
\node (016,2013Q2) at (6,4) {x};
\node (020,2013Q2) at (6,5) {x};
\node (022,2013Q2) at (6,6) {x};
\node (023,2013Q2) at (6,7) {x};
\node (024,2013Q2) at (6,8) {x};
\node (037,2013Q2) at (6,9) {x};
\node (038,2013Q2) at (6,10) {x};
\node (039,2013Q2) at (6,11) {x};
\node (048,2013Q2) at (6,12) {x};
\node (052,2013Q2) at (6,13) {x};
\node (085,2013Q2) at (6,14) {x};
\node (089,2013Q2) at (6,15) {x};
\node (095,2013Q2) at (6,16) {x};
\node (096,2013Q2) at (6,17) {x};
\node (098,2013Q2) at (6,18) {x};
\node (107,2013Q2) at (6,19) {x};
\node (110,2013Q2) at (6,20) {x};
\node (112,2013Q2) at (6,21) {x};

\node (004,2013Q3) at (7,1) {x};
\node (006,2013Q3) at (7,2) {x};
\node (015,2013Q3) at (7,3) {x};
\node (016,2013Q3) at (7,4) {x};
\node (020,2013Q3) at (7,5) {x};
\node (022,2013Q3) at (7,6) {x};
\node (023,2013Q3) at (7,7) {x};
\node (024,2013Q3) at (7,8) {x};
\node (037,2013Q3) at (7,9) {x};
\node (038,2013Q3) at (7,10) {x};
\node (039,2013Q3) at (7,11) {x};
\node (048,2013Q3) at (7,12) {x};
\node (052,2013Q3) at (7,13) {x};
\node (085,2013Q3) at (7,14) {x};
\node (089,2013Q3) at (7,15) {x};
\node (095,2013Q3) at (7,16) {x};
\node (096,2013Q3) at (7,17) {x};
\node (098,2013Q3) at (7,18) {x};
\node (107,2013Q3) at (7,19) {x};
\node (110,2013Q3) at (7,20) {x};
\node (112,2013Q3) at (7,21) {x};

\node (004,2013Q4) at (8,1) {x};
\node (006,2013Q4) at (8,2) {x};
\node (015,2013Q4) at (8,3) {x};
\node (016,2013Q4) at (8,4) {x};
\node (020,2013Q4) at (8,5) {x};
\node (022,2013Q4) at (8,6) {x};
\node (023,2013Q4) at (8,7) {x};
\node (024,2013Q4) at (8,8) {x};
\node (037,2013Q4) at (8,9) {x};
\node (038,2013Q4) at (8,10) {x};
\node (039,2013Q4) at (8,11) {x};
\node (048,2013Q4) at (8,12) {x};
\node (052,2013Q4) at (8,13) {x};
\node (085,2013Q4) at (8,14) {x};
\node (089,2013Q4) at (8,15) {x};
\node (095,2013Q4) at (8,16) {x};
\node (096,2013Q4) at (8,17) {x};
\node (098,2013Q4) at (8,18) {x};
\node (107,2013Q4) at (8,19) {x};
\node (110,2013Q4) at (8,20) {x};
\node (112,2013Q4) at (8,21) {x};

\node (004,2014Q1) at (9,1) {x};
\node (006,2014Q1) at (9,2) {x};
\node (015,2014Q1) at (9,3) {x};
\node (016,2014Q1) at (9,4) {x};
\node (022,2014Q1) at (9,6) {x};
\node (023,2014Q1) at (9,7) {x};
\node (024,2014Q1) at (9,8) {x};
\node (037,2014Q1) at (9,9) {x};
\node (038,2014Q1) at (9,10) {x};
\node (052,2014Q1) at (9,13) {x};
\node (085,2014Q1) at (9,14) {x};
\node (089,2014Q1) at (9,15) {x};
\node (095,2014Q1) at (9,16) {x};
\node (096,2014Q1) at (9,17) {x};
\node (098,2014Q1) at (9,18) {x};
\node (107,2014Q1) at (9,19) {x};
\node (110,2014Q1) at (9,20) {x};
\node (112,2014Q1) at (9,21) {x};

\node (004,2014Q2) at (10,1) {x};
\node (006,2014Q2) at (10,2) {x};
\node (015,2014Q2) at (10,3) {x};
\node (016,2014Q2) at (10,4) {x};
\node (020,2014Q2) at (10,5) {x};
\node (022,2014Q2) at (10,6) {x};
\node (023,2014Q2) at (10,7) {x};
\node (024,2014Q2) at (10,8) {x};
\node (037,2014Q2) at (10,9) {x};
\node (038,2014Q2) at (10,10) {x};
\node (039,2014Q2) at (10,11) {x};
\node (048,2014Q2) at (10,12) {x};
\node (052,2014Q2) at (10,13) {x};
\node (085,2014Q2) at (10,14) {x};
\node (089,2014Q2) at (10,15) {x};
\node (095,2014Q2) at (10,16) {x};
\node (096,2014Q2) at (10,17) {x};
\node (098,2014Q2) at (10,18) {x};
\node (107,2014Q2) at (10,19) {x};
\node (110,2014Q2) at (10,20) {x};
\node (112,2014Q2) at (10,21) {x};

\node (004,2014Q3) at (11,1) {x};
\node (006,2014Q3) at (11,2) {x};
\node (015,2014Q3) at (11,3) {x};
\node (016,2014Q3) at (11,4) {x};
\node (020,2014Q3) at (11,5) {x};
\node (022,2014Q3) at (11,6) {x};
\node (023,2014Q3) at (11,7) {x};
\node (024,2014Q3) at (11,8) {x};
\node (037,2014Q3) at (11,9) {x};
\node (038,2014Q3) at (11,10) {x};
\node (039,2014Q3) at (11,11) {x};
\node (048,2014Q3) at (11,12) {x};
\node (052,2014Q3) at (11,13) {x};
\node (085,2014Q3) at (11,14) {x};
\node (089,2014Q3) at (11,15) {x};
\node (095,2014Q3) at (11,16) {x};
\node (096,2014Q3) at (11,17) {x};
\node (098,2014Q3) at (11,18) {x};
\node (107,2014Q3) at (11,19) {x};
\node (110,2014Q3) at (11,20) {x};
\node (112,2014Q3) at (11,21) {x};

\node (004,2014Q4) at (12,1) {x};
\node (006,2014Q4) at (12,2) {x};
\node (015,2014Q4) at (12,3) {x};
\node (016,2014Q4) at (12,4) {x};
\node (020,2014Q4) at (12,5) {x};
\node (022,2014Q4) at (12,6) {x};
\node (023,2014Q4) at (12,7) {x};
\node (024,2014Q4) at (12,8) {x};
\node (037,2014Q4) at (12,9) {x};
\node (038,2014Q4) at (12,10) {x};
\node (039,2014Q4) at (12,11) {x};
\node (048,2014Q4) at (12,12) {x};
\node (052,2014Q4) at (12,13) {x};
\node (085,2014Q4) at (12,14) {x};
\node (089,2014Q4) at (12,15) {x};
\node (095,2014Q4) at (12,16) {x};
\node (096,2014Q4) at (12,17) {x};
\node (098,2014Q4) at (12,18) {x};
\node (107,2014Q4) at (12,19) {x};
\node (110,2014Q4) at (12,20) {x};
\node (112,2014Q4) at (12,21) {x};

\node (004,2015Q1) at (13,1) {x};
\node (006,2015Q1) at (13,2) {x};
\node (015,2015Q1) at (13,3) {x};
\node (016,2015Q1) at (13,4) {x};
\node (020,2015Q1) at (13,5) {x};
\node (022,2015Q1) at (13,6) {x};
\node (023,2015Q1) at (13,7) {x};
\node (024,2015Q1) at (13,8) {x};
\node (037,2015Q1) at (13,9) {x};
\node (038,2015Q1) at (13,10) {x};
\node (039,2015Q1) at (13,11) {x};
\node (048,2015Q1) at (13,12) {x};
\node (052,2015Q1) at (13,13) {x};
\node (085,2015Q1) at (13,14) {x};
\node (089,2015Q1) at (13,15) {x};
\node (095,2015Q1) at (13,16) {x};
\node (096,2015Q1) at (13,17) {x};
\node (098,2015Q1) at (13,18) {x};
\node (110,2015Q1) at (13,20) {x};
\node (112,2015Q1) at (13,21) {x};

\node (006,2015Q2) at (14,2) {x};
\node (015,2015Q2) at (14,3) {x};
\node (016,2015Q2) at (14,4) {x};
\node (020,2015Q2) at (14,5) {x};
\node (023,2015Q2) at (14,7) {x};
\node (024,2015Q2) at (14,8) {x};
\node (037,2015Q2) at (14,9) {x};
\node (038,2015Q2) at (14,10) {x};
\node (039,2015Q2) at (14,11) {x};
\node (048,2015Q2) at (14,12) {x};
\node (052,2015Q2) at (14,13) {x};
\node (085,2015Q2) at (14,14) {x};
\node (089,2015Q2) at (14,15) {x};
\node (095,2015Q2) at (14,16) {x};
\node (096,2015Q2) at (14,17) {x};
\node (098,2015Q2) at (14,18) {x};
\node (107,2015Q2) at (14,19) {x};
\node (110,2015Q2) at (14,20) {x};
\node (112,2015Q2) at (14,21) {x};

\node (004,2015Q3) at (15,1) {x};
\node (006,2015Q3) at (15,2) {x};
\node (015,2015Q3) at (15,3) {x};
\node (016,2015Q3) at (15,4) {x};
\node (020,2015Q3) at (15,5) {x};
\node (022,2015Q3) at (15,6) {x};
\node (023,2015Q3) at (15,7) {x};
\node (024,2015Q3) at (15,8) {x};
\node (037,2015Q3) at (15,9) {x};
\node (039,2015Q3) at (15,11) {x};
\node (048,2015Q3) at (15,12) {x};
\node (052,2015Q3) at (15,13) {x};
\node (085,2015Q3) at (15,14) {x};
\node (089,2015Q3) at (15,15) {x};
\node (095,2015Q3) at (15,16) {x};
\node (096,2015Q3) at (15,17) {x};
\node (098,2015Q3) at (15,18) {x};
\node (107,2015Q3) at (15,19) {x};
\node (112,2015Q3) at (15,21) {x};

\node (004,2015Q4) at (16,1) {x};
\node (006,2015Q4) at (16,2) {x};
\node (015,2015Q4) at (16,3) {x};
\node (016,2015Q4) at (16,4) {x};
\node (020,2015Q4) at (16,5) {x};
\node (023,2015Q4) at (16,7) {x};
\node (024,2015Q4) at (16,8) {x};
\node (037,2015Q4) at (16,9) {x};
\node (038,2015Q4) at (16,10) {x};
\node (039,2015Q4) at (16,11) {x};
\node (052,2015Q4) at (16,13) {x};
\node (085,2015Q4) at (16,14) {x};
\node (089,2015Q4) at (16,15) {x};
\node (095,2015Q4) at (16,16) {x};
\node (096,2015Q4) at (16,17) {x};
\node (098,2015Q4) at (16,18) {x};
\node (107,2015Q4) at (16,19) {x};
\node (110,2015Q4) at (16,20) {x};
\node (112,2015Q4) at (16,21) {x};

\node (004,2016Q1) at (17,1) {x};
\node (006,2016Q1) at (17,2) {x};
\node (015,2016Q1) at (17,3) {x};
\node (016,2016Q1) at (17,4) {x};
\node (020,2016Q1) at (17,5) {x};
\node (022,2016Q1) at (17,6) {x};
\node (023,2016Q1) at (17,7) {x};
\node (024,2016Q1) at (17,8) {x};
\node (037,2016Q1) at (17,9) {x};
\node (038,2016Q1) at (17,10) {x};
\node (039,2016Q1) at (17,11) {x};
\node (048,2016Q1) at (17,12) {x};
\node (052,2016Q1) at (17,13) {x};
\node (085,2016Q1) at (17,14) {x};
\node (089,2016Q1) at (17,15) {x};
\node (095,2016Q1) at (17,16) {x};
\node (096,2016Q1) at (17,17) {x};
\node (098,2016Q1) at (17,18) {x};
\node (110,2016Q1) at (17,20) {x};
\node (112,2016Q1) at (17,21) {x};

\node (004,2016Q2) at (18,1) {x};
\node (006,2016Q2) at (18,2) {x};
\node (015,2016Q2) at (18,3) {x};
\node (016,2016Q2) at (18,4) {x};
\node (020,2016Q2) at (18,5) {x};
\node (022,2016Q2) at (18,6) {x};
\node (023,2016Q2) at (18,7) {x};
\node (024,2016Q2) at (18,8) {x};
\node (037,2016Q2) at (18,9) {x};
\node (038,2016Q2) at (18,10) {x};
\node (039,2016Q2) at (18,11) {x};
\node (048,2016Q2) at (18,12) {x};
\node (052,2016Q2) at (18,13) {x};
\node (085,2016Q2) at (18,14) {x};
\node (089,2016Q2) at (18,15) {x};
\node (095,2016Q2) at (18,16) {x};
\node (096,2016Q2) at (18,17) {x};
\node (107,2016Q2) at (18,19) {x};
\node (110,2016Q2) at (18,20) {x};
\node (112,2016Q2) at (18,21) {x};

\node (006,2016Q3) at (19,2) {x};
\node (015,2016Q3) at (19,3) {x};
\node (016,2016Q3) at (19,4) {x};
\node (020,2016Q3) at (19,5) {x};
\node (022,2016Q3) at (19,6) {x};
\node (023,2016Q3) at (19,7) {x};
\node (024,2016Q3) at (19,8) {x};
\node (037,2016Q3) at (19,9) {x};
\node (038,2016Q3) at (19,10) {x};
\node (039,2016Q3) at (19,11) {x};
\node (048,2016Q3) at (19,12) {x};
\node (052,2016Q3) at (19,13) {x};
\node (085,2016Q3) at (19,14) {x};
\node (089,2016Q3) at (19,15) {x};
\node (095,2016Q3) at (19,16) {x};
\node (096,2016Q3) at (19,17) {x};
\node (098,2016Q3) at (19,18) {x};
\node (110,2016Q3) at (19,20) {x};
\node (112,2016Q3) at (19,21) {x};

\node (004,2016Q4) at (20,1) {x};
\node (006,2016Q4) at (20,2) {x};
\node (015,2016Q4) at (20,3) {x};
\node (016,2016Q4) at (20,4) {x};
\node (020,2016Q4) at (20,5) {x};
\node (023,2016Q4) at (20,7) {x};
\node (024,2016Q4) at (20,8) {x};
\node (037,2016Q4) at (20,9) {x};
\node (038,2016Q4) at (20,10) {x};
\node (039,2016Q4) at (20,11) {x};
\node (048,2016Q4) at (20,12) {x};
\node (052,2016Q4) at (20,13) {x};
\node (085,2016Q4) at (20,14) {x};
\node (089,2016Q4) at (20,15) {x};
\node (095,2016Q4) at (20,16) {x};
\node (096,2016Q4) at (20,17) {x};
\node (098,2016Q4) at (20,18) {x};
\node (107,2016Q4) at (20,19) {x};
\node (110,2016Q4) at (20,20) {x};
\node (112,2016Q4) at (20,21) {x};

\node (004,2017Q1) at (21,1) {x};
\node (006,2017Q1) at (21,2) {x};
\node (015,2017Q1) at (21,3) {x};
\node (016,2017Q1) at (21,4) {x};
\node (020,2017Q1) at (21,5) {x};
\node (022,2017Q1) at (21,6) {x};
\node (023,2017Q1) at (21,7) {x};
\node (024,2017Q1) at (21,8) {x};
\node (037,2017Q1) at (21,9) {x};
\node (038,2017Q1) at (21,10) {x};
\node (039,2017Q1) at (21,11) {x};
\node (048,2017Q1) at (21,12) {x};
\node (085,2017Q1) at (21,14) {x};
\node (089,2017Q1) at (21,15) {x};
\node (095,2017Q1) at (21,16) {x};
\node (096,2017Q1) at (21,17) {x};
\node (098,2017Q1) at (21,18) {x};
\node (110,2017Q1) at (21,20) {x};
\node (112,2017Q1) at (21,21) {x};

\node (004,2017Q2) at (22,1) {x};
\node (006,2017Q2) at (22,2) {x};
\node (015,2017Q2) at (22,3) {x};
\node (016,2017Q2) at (22,4) {x};
\node (020,2017Q2) at (22,5) {x};
\node (022,2017Q2) at (22,6) {x};
\node (023,2017Q2) at (22,7) {x};
\node (024,2017Q2) at (22,8) {x};
\node (037,2017Q2) at (22,9) {x};
\node (038,2017Q2) at (22,10) {x};
\node (039,2017Q2) at (22,11) {x};
\node (048,2017Q2) at (22,12) {x};
\node (052,2017Q2) at (22,13) {x};
\node (085,2017Q2) at (22,14) {x};
\node (089,2017Q2) at (22,15) {x};
\node (095,2017Q2) at (22,16) {x};
\node (096,2017Q2) at (22,17) {x};
\node (098,2017Q2) at (22,18) {x};
\node (107,2017Q2) at (22,19) {x};
\node (110,2017Q2) at (22,20) {x};
\node (112,2017Q2) at (22,21) {x};

\node (004,2017Q3) at (23,1) {x};
\node (006,2017Q3) at (23,2) {x};
\node (015,2017Q3) at (23,3) {x};
\node (016,2017Q3) at (23,4) {x};
\node (020,2017Q3) at (23,5) {x};
\node (022,2017Q3) at (23,6) {x};
\node (023,2017Q3) at (23,7) {x};
\node (024,2017Q3) at (23,8) {x};
\node (037,2017Q3) at (23,9) {x};
\node (038,2017Q3) at (23,10) {x};
\node (039,2017Q3) at (23,11) {x};
\node (048,2017Q3) at (23,12) {x};
\node (085,2017Q3) at (23,14) {x};
\node (089,2017Q3) at (23,15) {x};
\node (095,2017Q3) at (23,16) {x};
\node (096,2017Q3) at (23,17) {x};
\node (098,2017Q3) at (23,18) {x};
\node (107,2017Q3) at (23,19) {x};
\node (110,2017Q3) at (23,20) {x};
\node (112,2017Q3) at (23,21) {x};

\node (004,2017Q4) at (24,1) {x};
\node (006,2017Q4) at (24,2) {x};
\node (015,2017Q4) at (24,3) {x};
\node (016,2017Q4) at (24,4) {x};
\node (020,2017Q4) at (24,5) {x};
\node (022,2017Q4) at (24,6) {x};
\node (023,2017Q4) at (24,7) {x};
\node (024,2017Q4) at (24,8) {x};
\node (037,2017Q4) at (24,9) {x};
\node (038,2017Q4) at (24,10) {x};
\node (039,2017Q4) at (24,11) {x};
\node (048,2017Q4) at (24,12) {x};
\node (052,2017Q4) at (24,13) {x};
\node (085,2017Q4) at (24,14) {x};
\node (089,2017Q4) at (24,15) {x};
\node (095,2017Q4) at (24,16) {x};
\node (096,2017Q4) at (24,17) {x};
\node (098,2017Q4) at (24,18) {x};
\node (110,2017Q4) at (24,20) {x};
\node (112,2017Q4) at (24,21) {x};

\node (004,2018Q1) at (25,1) {x};
\node (006,2018Q1) at (25,2) {x};
\node (015,2018Q1) at (25,3) {x};
\node (016,2018Q1) at (25,4) {x};
\node (020,2018Q1) at (25,5) {x};
\node (022,2018Q1) at (25,6) {x};
\node (023,2018Q1) at (25,7) {x};
\node (024,2018Q1) at (25,8) {x};
\node (037,2018Q1) at (25,9) {x};
\node (038,2018Q1) at (25,10) {x};
\node (039,2018Q1) at (25,11) {x};
\node (048,2018Q1) at (25,12) {x};
\node (052,2018Q1) at (25,13) {x};
\node (085,2018Q1) at (25,14) {x};
\node (089,2018Q1) at (25,15) {x};
\node (095,2018Q1) at (25,16) {x};
\node (096,2018Q1) at (25,17) {x};
\node (098,2018Q1) at (25,18) {x};
\node (107,2018Q1) at (25,19) {x};
\node (110,2018Q1) at (25,20) {x};
\node (112,2018Q1) at (25,21) {x};

\node (004,2018Q2) at (26,1) {x};
\node (006,2018Q2) at (26,2) {x};
\node (015,2018Q2) at (26,3) {x};
\node (016,2018Q2) at (26,4) {x};
\node (022,2018Q2) at (26,6) {x};
\node (023,2018Q2) at (26,7) {x};
\node (024,2018Q2) at (26,8) {x};
\node (037,2018Q2) at (26,9) {x};
\node (038,2018Q2) at (26,10) {x};
\node (039,2018Q2) at (26,11) {x};
\node (048,2018Q2) at (26,12) {x};
\node (052,2018Q2) at (26,13) {x};
\node (085,2018Q2) at (26,14) {x};
\node (089,2018Q2) at (26,15) {x};
\node (095,2018Q2) at (26,16) {x};
\node (096,2018Q2) at (26,17) {x};
\node (098,2018Q2) at (26,18) {x};
\node (107,2018Q2) at (26,19) {x};
\node (110,2018Q2) at (26,20) {x};
\node (112,2018Q2) at (26,21) {x};

\node (006,2018Q3) at (27,2) {x};
\node (015,2018Q3) at (27,3) {x};
\node (016,2018Q3) at (27,4) {x};
\node (020,2018Q3) at (27,5) {x};
\node (022,2018Q3) at (27,6) {x};
\node (023,2018Q3) at (27,7) {x};
\node (024,2018Q3) at (27,8) {x};
\node (037,2018Q3) at (27,9) {x};
\node (038,2018Q3) at (27,10) {x};
\node (039,2018Q3) at (27,11) {x};
\node (048,2018Q3) at (27,12) {x};
\node (052,2018Q3) at (27,13) {x};
\node (085,2018Q3) at (27,14) {x};
\node (089,2018Q3) at (27,15) {x};
\node (095,2018Q3) at (27,16) {x};
\node (096,2018Q3) at (27,17) {x};
\node (098,2018Q3) at (27,18) {x};
\node (107,2018Q3) at (27,19) {x};
\node (110,2018Q3) at (27,20) {x};
\node (112,2018Q3) at (27,21) {x};

\node (004,2018Q4) at (28,1) {x};
\node (006,2018Q4) at (28,2) {x};
\node (015,2018Q4) at (28,3) {x};
\node (016,2018Q4) at (28,4) {x};
\node (020,2018Q4) at (28,5) {x};
\node (022,2018Q4) at (28,6) {x};
\node (023,2018Q4) at (28,7) {x};
\node (024,2018Q4) at (28,8) {x};
\node (037,2018Q4) at (28,9) {x};
\node (038,2018Q4) at (28,10) {x};
\node (039,2018Q4) at (28,11) {x};
\node (048,2018Q4) at (28,12) {x};
\node (085,2018Q4) at (28,14) {x};
\node (089,2018Q4) at (28,15) {x};
\node (095,2018Q4) at (28,16) {x};
\node (096,2018Q4) at (28,17) {x};
\node (098,2018Q4) at (28,18) {x};
\node (107,2018Q4) at (28,19) {x};
\node (110,2018Q4) at (28,20) {x};
\node (112,2018Q4) at (28,21) {x};

\node (004,2019Q1) at (29,1) {x};
\node (006,2019Q1) at (29,2) {x};
\node (015,2019Q1) at (29,3) {x};
\node (016,2019Q1) at (29,4) {x};
\node (020,2019Q1) at (29,5) {x};
\node (022,2019Q1) at (29,6) {x};
\node (023,2019Q1) at (29,7) {x};
\node (024,2019Q1) at (29,8) {x};
\node (037,2019Q1) at (29,9) {x};
\node (038,2019Q1) at (29,10) {x};
\node (039,2019Q1) at (29,11) {x};
\node (048,2019Q1) at (29,12) {x};
\node (052,2019Q1) at (29,13) {x};
\node (085,2019Q1) at (29,14) {x};
\node (089,2019Q1) at (29,15) {x};
\node (095,2019Q1) at (29,16) {x};
\node (096,2019Q1) at (29,17) {x};
\node (098,2019Q1) at (29,18) {x};
\node (107,2019Q1) at (29,19) {x};
\node (110,2019Q1) at (29,20) {x};
\node (112,2019Q1) at (29,21) {x};

\node (004,2019Q2) at (30,1) {x};
\node (006,2019Q2) at (30,2) {x};
\node (015,2019Q2) at (30,3) {x};
\node (016,2019Q2) at (30,4) {x};
\node (020,2019Q2) at (30,5) {x};
\node (022,2019Q2) at (30,6) {x};
\node (023,2019Q2) at (30,7) {x};
\node (024,2019Q2) at (30,8) {x};
\node (037,2019Q2) at (30,9) {x};
\node (038,2019Q2) at (30,10) {x};
\node (039,2019Q2) at (30,11) {x};
\node (048,2019Q2) at (30,12) {x};
\node (085,2019Q2) at (30,14) {x};
\node (089,2019Q2) at (30,15) {x};
\node (095,2019Q2) at (30,16) {x};
\node (096,2019Q2) at (30,17) {x};
\node (098,2019Q2) at (30,18) {x};
\node (110,2019Q2) at (30,20) {x};
\node (112,2019Q2) at (30,21) {x};

\node (004,2019Q3) at (31,1) {x};
\node (006,2019Q3) at (31,2) {x};
\node (015,2019Q3) at (31,3) {x};
\node (016,2019Q3) at (31,4) {x};
\node (020,2019Q3) at (31,5) {x};
\node (022,2019Q3) at (31,6) {x};
\node (023,2019Q3) at (31,7) {x};
\node (024,2019Q3) at (31,8) {x};
\node (037,2019Q3) at (31,9) {x};
\node (038,2019Q3) at (31,10) {x};
\node (039,2019Q3) at (31,11) {x};
\node (048,2019Q3) at (31,12) {x};
\node (052,2019Q3) at (31,13) {x};
\node (085,2019Q3) at (31,14) {x};
\node (089,2019Q3) at (31,15) {x};
\node (095,2019Q3) at (31,16) {x};
\node (096,2019Q3) at (31,17) {x};
\node (098,2019Q3) at (31,18) {x};
\node (107,2019Q3) at (31,19) {x};
\node (110,2019Q3) at (31,20) {x};
\node (112,2019Q3) at (31,21) {x};

\node (004,2019Q4) at (32,1) {x};
\node (006,2019Q4) at (32,2) {x};
\node (015,2019Q4) at (32,3) {x};
\node (016,2019Q4) at (32,4) {x};
\node (020,2019Q4) at (32,5) {x};
\node (022,2019Q4) at (32,6) {x};
\node (023,2019Q4) at (32,7) {x};
\node (024,2019Q4) at (32,8) {x};
\node (037,2019Q4) at (32,9) {x};
\node (038,2019Q4) at (32,10) {x};
\node (039,2019Q4) at (32,11) {x};
\node (048,2019Q4) at (32,12) {x};
\node (085,2019Q4) at (32,14) {x};
\node (089,2019Q4) at (32,15) {x};
\node (095,2019Q4) at (32,16) {x};
\node (096,2019Q4) at (32,17) {x};
\node (098,2019Q4) at (32,18) {x};
\node (107,2019Q4) at (32,19) {x};
\node (110,2019Q4) at (32,20) {x};
\node (112,2019Q4) at (32,21) {x};

\node (006,2020Q1) at (33,2) {x};
\node (015,2020Q1) at (33,3) {x};
\node (016,2020Q1) at (33,4) {x};
\node (020,2020Q1) at (33,5) {x};
\node (022,2020Q1) at (33,6) {x};
\node (023,2020Q1) at (33,7) {x};
\node (024,2020Q1) at (33,8) {x};
\node (038,2020Q1) at (33,10) {x};
\node (039,2020Q1) at (33,11) {x};
\node (048,2020Q1) at (33,12) {x};
\node (052,2020Q1) at (33,13) {x};
\node (085,2020Q1) at (33,14) {x};
\node (089,2020Q1) at (33,15) {x};
\node (095,2020Q1) at (33,16) {x};
\node (096,2020Q1) at (33,17) {x};
\node (098,2020Q1) at (33,18) {x};
\node (107,2020Q1) at (33,19) {x};
\node (110,2020Q1) at (33,20) {x};
\node (112,2020Q1) at (33,21) {x};

\node (004,2020Q2) at (34,1) {x};
\node (006,2020Q2) at (34,2) {x};
\node (015,2020Q2) at (34,3) {x};
\node (016,2020Q2) at (34,4) {x};
\node (020,2020Q2) at (34,5) {x};
\node (022,2020Q2) at (34,6) {x};
\node (023,2020Q2) at (34,7) {x};
\node (024,2020Q2) at (34,8) {x};
\node (037,2020Q2) at (34,9) {x};
\node (038,2020Q2) at (34,10) {x};
\node (039,2020Q2) at (34,11) {x};
\node (048,2020Q2) at (34,12) {x};
\node (085,2020Q2) at (34,14) {x};
\node (089,2020Q2) at (34,15) {x};
\node (095,2020Q2) at (34,16) {x};
\node (096,2020Q2) at (34,17) {x};
\node (098,2020Q2) at (34,18) {x};
\node (107,2020Q2) at (34,19) {x};
\node (110,2020Q2) at (34,20) {x};
\node (112,2020Q2) at (34,21) {x};

\node (004,2020Q2) at (35,1) {x};
\node (006,2020Q2) at (35,2) {x};
\node (015,2020Q2) at (35,3) {x};
\node (016,2020Q2) at (35,4) {x};
\node (020,2020Q2) at (35,5) {x};
\node (022,2020Q2) at (35,6) {x};
\node (023,2020Q2) at (35,7) {x};
\node (024,2020Q2) at (35,8) {x};
\node (037,2020Q2) at (35,9) {x};
\node (038,2020Q2) at (35,10) {x};
\node (039,2020Q2) at (35,11) {x};
\node (048,2020Q2) at (35,12) {x};
\node (048,2020Q2) at (35,13) {x};
\node (085,2020Q2) at (35,14) {x};
\node (089,2020Q2) at (35,15) {x};
\node (095,2020Q2) at (35,16) {x};
\node (096,2020Q2) at (35,17) {x};
\node (098,2020Q2) at (35,18) {x};
\node (107,2020Q2) at (35,19) {x};
\node (110,2020Q2) at (35,20) {x};
\node (112,2020Q2) at (35,21) {X};
\end{tikzpicture}
{\footnotesize \\ Notes: A slot is marked with the notation x if a forecast is provided but left blank otherwise.}
\end{figure}
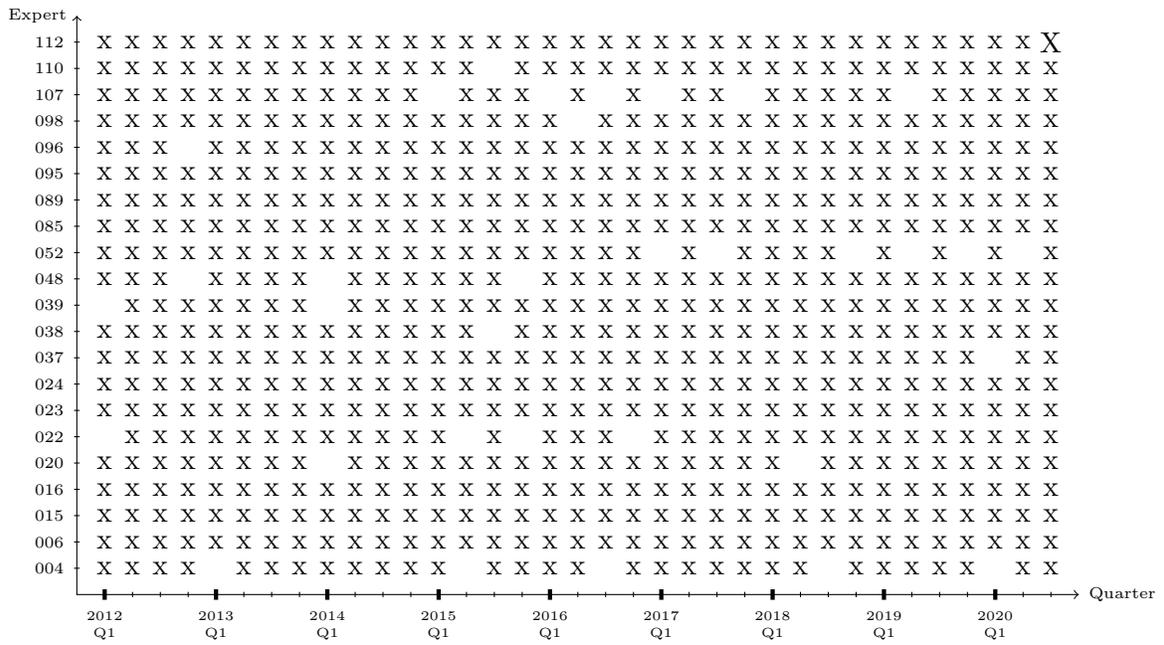 

%% file: Figure2_new.tex
\begin{figure}[t]
\caption{Sample Correlation Coefficients of Forecast Errors}\label{Figure2_new}
\centering
\vspace{0.3cm}
\includegraphics[scale=0.7]{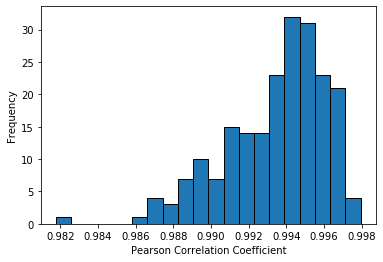}
\end{figure}

%% file: Figure3_new.tex
\begin{figure}[t]
\caption{The Frequency of Number of Experts in the Committee Performing Best per Round}
\label{Figure3_new}
\centering
\vspace{0.3cm}
\subfloat[HECA without delayed announcements (Algorithm~\ref{HECA} with $l=2$)]
{\includegraphics[scale=0.7]{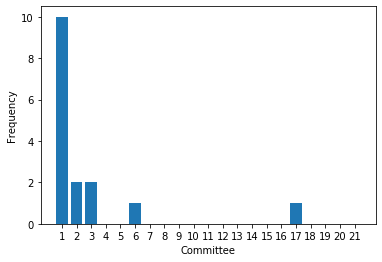}}
\qquad
\subfloat[HECA with delayed announcements (Algorithm~\ref{HECA_delayed_announcements} with $l=1$)]
{\includegraphics[scale=0.7]{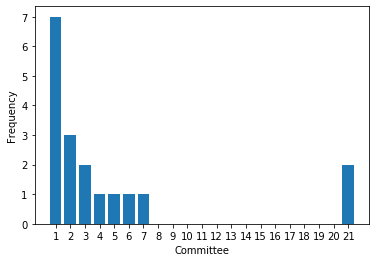}}
\end{figure} 

%% file: Table1_new.tex
\begin{table}[t]
\caption{Sample Variances of Forecast Errors}
\label{Table1_new}
\centering
\begin{tabular}{ccccccc}
\toprule
8.065 & 8.687 & 8.533 & 8.211 & 8.139 & 8.405 & 8.139 \\
8.178 & 8.167 & 8.454 & 8.310 & 8.411 & 8.246 & 7.935 \\
8.337 & 8.482 & 8.532 & 8.397 & 8.265 & 8.882 & 8.260
\end{tabular}
\end{table} 

%% file: Table2_new.tex
\begin{table}[t]
\caption{Forecasting Loss of the Equal-Weight Scheme and HECA}
\label{Table2_new}
\centering
\begin{tabular}{rrrrrr}
\toprule
& \multicolumn{1}{c}{(1) Equal-Weight} & \multicolumn{1}{c}{(2) HECA} & \multicolumn{1}{c}{(3) HECA} & \multicolumn{1}{c}{Difference} & \multicolumn{1}{c}{Difference}\\
& \multicolumn{1}{c}{Scheme} & \multicolumn{1}{c}{Algorithm~\ref{HECA}}  & \multicolumn{1}{c}{Algorithm~\ref{HECA_delayed_announcements}} & \multicolumn{1}{c}{(1) - (2)} & \multicolumn{1}{c}{(1) - (3)}\\
\cmidrule(r){1-6}
2016Q2                   & 0.0060   & -\hspace{0.5cm} & 0.0003   & -\hspace{0.5cm} & 0.0057\\
2016Q3                   & 0.0060   & -\hspace{0.5cm} & 0.0053   & -\hspace{0.5cm} & 0.0007\\
2016Q4                   & 0.0051   & 0.0003          & 0.0023   & 0.0048          & 0.0028\\
\textcolor{blue}{2017Q1} & 0.1630   & 0.1000          & 0.2629   & 0.0630          & -0.0999\\
2017Q2                   & 0.6170   & 0.7911          & 0.5335   & -0.1741         & 0.0835\\
2017Q3                   & 0.8522   & 0.9681          & 0.7981   & -0.1160         & 0.0540\\
2017Q4                   & 1.0559   & 1.0032          & 0.9484   & 0.0526          & 0.1074\\
\textcolor{blue}{2018Q1} & 0.4457   & 0.3646          & 0.2621   & 0.0812          & 0.1837\\
2018Q2                   & 0.0393   & 0.0476          & 0.0926   & -0.0083         & -0.0532\\
2018Q3                   & 0.2743   & 0.3592          & 0.4149   & -0.0849         & -0.1406\\
2018Q4                   & 0.9174   & 0.9614          & 0.9729   & -0.0441         & -0.0556\\
\textcolor{blue}{2019Q1} & 0.5464   & 0.6747          & 0.6253   & -0.1284         & -0.0790\\
2019Q2                   & 0.5781   & 0.4997          & 0.4996   & 0.0784          & 0.0785\\
2019Q3                   & 0.1249   & 0.1667          & 0.1241   & -0.0418         & 0.0008\\
2019Q4                   & 0.1521   & 0.0827          & 0.0905   & 0.0694          & 0.0617\\
\textcolor{blue}{2020Q1} & 19.4562  & 19.3347         & 19.3059  & 0.1215          & 0.1502\\
2020Q2                   & 250.3428 & 248.8660        & 244.8226 & 1.4769          & 5.5202\\
2020Q3                   & 29.2636  & 28.7480         & 27.9170  & 0.5156          & 1.3465\\
\end{tabular}
\end{table} 

%% file: Table3_new.tex
\begin{table}[t]
\caption{Cumulative Forecasting Loss and Average Regret of HECA}
\label{Table3_new}
\centering
\begin{tabular}{rrrr}
\toprule
\multicolumn{1}{l}{HECA} & & & \\[0.1cm]
& \multicolumn{1}{r}{Algorithm~\ref{HECA}} & \multicolumn{1}{r}{Best Committee} & \multicolumn{1}{r}{Difference} \\
\cmidrule(r){1-4}
2016Q4                   & 0.0003          & 0.0020          & -0.0017   \\
\textcolor{blue}{2017Q1} & 0.1003          & 0.0581          & 0.0422    \\
2017Q2                   & 0.8914          & 0.7823          & 0.1091    \\
2017Q3                   & 1.8595          & 1.7421          & 0.1174    \\
2017Q4                   & 2.8628          & 2.7036          & 0.1592    \\
\textcolor{blue}{2018Q1} & 3.2274          & 3.2053          & 0.0220    \\
2018Q2                   & 3.2749          & 3.3391          & -0.0642   \\
2018Q3                   & 3.6341          & 3.8457          & -0.2116   \\
2018Q4                   & 4.5956          & 4.5637          & 0.0319    \\
\textcolor{blue}{2019Q1} & 5.2703          & 5.2582          & 0.0121    \\
2019Q2                   & 5.7700          & 5.6740          & 0.0960    \\
2019Q3                   & 5.9367          & 5.6895          & 0.2472    \\
2019Q4                   & 6.0194          & 5.7567          & 0.2627    \\
\textcolor{blue}{2020Q1} & 25.3541         & 25.2129         & 0.1412    \\
2020Q2                   & 274.2201        & 267.6883        & 6.5317    \\
2020Q3                   & 302.9680        & 295.6054        & 7.3627    \\
\cmidrule(r){1-4}
\multicolumn{2}{l}{\textcolor{blue}{Average Regret} (2016Q4 $\sim$ 2020Q3)} & & 0.4602    \\[0.6cm]
\multicolumn{3}{l}{HECA with Delayed Announcements} & \\[0.1cm]
& \multicolumn{1}{r}{Algorithm~\ref{HECA_delayed_announcements}} & \multicolumn{1}{r}{Best Committee} & \multicolumn{1}{r}{Difference} \\
\cmidrule(r){1-4}
2016Q2                   & 0.0003          & 0.0164          & -0.0161   \\
2016Q3                   & 0.0056          & 0.0233          & -0.0176   \\
2016Q4                   & 0.0079          & 0.0252          & -0.0173   \\
\textcolor{blue}{2017Q1} & 0.2708          & 0.0814          & 0.1894    \\
2017Q2                   & 0.8043          & 0.8056          & -0.0013   \\
2017Q3                   & 1.6024          & 1.4135          & 0.1889    \\
2017Q4                   & 2.5509          & 2.3750          & 0.1795    \\
\textcolor{blue}{2018Q1} & 2.8129          & 2.5417          & 0.2712    \\
2018Q2                   & 2.9055          & 2.6755          & 0.2300    \\
2018Q3                   & 3.3204          & 3.1821          & 0.1383    \\
2018Q4                   & 4.2933          & 4.3225          & -0.0292   \\
\textcolor{blue}{2019Q1} & 4.9186          & 4.7022          & 0.2165    \\
2019Q2                   & 5.4183          & 5.1179          & 0.3003    \\
2019Q3                   & 5.5424          & 5.1334          & 0.4090    \\
2019Q4                   & 5.6328          & 5.2006          & 0.4322    \\
\textcolor{blue}{2020Q1} & 24.9388         & 24.6568         & 0.2820    \\
2020Q2                   & 269.7614        & 264.0279        & 5.7335    \\
2020Q3                   & 297.6784        & 291.9450        & 5.7335    \\
\cmidrule(r){1-4}
\multicolumn{2}{l}{\textcolor{blue}{Average Regret} (2016Q2 $\sim$ 2020Q3)} & & 0.3185
\end{tabular}
\end{table} 

%% file: Table4_new.tex
\begin{table}[t]
\caption{Forecasting Loss of the Equal-Weight Scheme, HECA, and Fictitious Play}
\label{Table4_new}
\centering
\begin{tabular}{rrrrrr}
\toprule
& \multicolumn{1}{c}{Equal-Weight} & \multicolumn{1}{c}{HECA} & \multicolumn{1}{c}{Fictitious} & \multicolumn{1}{c}{HECA} & \multicolumn{1}{c}{Fictitious}\\
& \multicolumn{1}{c}{Scheme} & \multicolumn{1}{c}{Algorithm~\ref{HECA}}  & \multicolumn{1}{c}{Play~(\ref{EFP-ECA})} & \multicolumn{1}{c}{Algorithm~\ref{HECA_delayed_announcements}} & \multicolumn{1}{c}{Play~(\ref{EFP-ECA2})}\\
\cmidrule(r){1-6}
2016Q2                   & 0.0060   & -\hspace{0.5cm} & -\hspace{0.5cm} & 0.0003   & 0.0003\\
2016Q3                   & 0.0060   & -\hspace{0.5cm} & -\hspace{0.5cm} & 0.0053   & 0.0053\\
2016Q4                   & 0.0051   & 0.0003          & 0.0003          & 0.0023   & 0.0023\\
\textcolor{blue}{2017Q1} & 0.1630   & 0.1000          & 0.1000          & 0.2629   & 0.2629\\
2017Q2                   & 0.6170   & 0.7911          & 0.7911          & 0.5335   & 0.5331\\
2017Q3                   & 0.8522   & 0.9681          & 0.9681          & 0.7981   & 0.7981\\
2017Q4                   & 1.0559   & 1.0032          & 1.0033          & 0.9484   & 0.9484\\
\textcolor{blue}{2018Q1} & 0.4457   & 0.3646          & 0.3645          & 0.2621   & 0.2622\\
2018Q2                   & 0.0393   & 0.0476          & 0.0476          & 0.0926   & 0.0925\\
2018Q3                   & 0.2743   & 0.3592          & 0.3594          & 0.4149   & 0.4148\\
2018Q4                   & 0.9174   & 0.9614          & 0.9618          & 0.9729   & 0.9731\\
\textcolor{blue}{2019Q1} & 0.5464   & 0.6747          & 0.6751          & 0.6253   & 0.6253\\
2019Q2                   & 0.5781   & 0.4997          & 0.5000          & 0.4996   & 0.4996\\
2019Q3                   & 0.1249   & 0.1667          & 0.1658          & 0.1241   & 0.1244\\
2019Q4                   & 0.1521   & 0.0827          & 0.0828          & 0.0905   & 0.0905\\
\textcolor{blue}{2020Q1} & 19.4562  & 19.3347         & 19.3325         & 19.3059  & 19.3055\\
2020Q2                   & 250.3428 & 248.8660        & 248.9044        & 244.8226 & 244.8175\\
2020Q3                   & 29.2636  & 28.7480         & 28.7447         & 27.9170  & 27.9170\\
\end{tabular}
\end{table} 

%% file: Table5_new.tex
\begin{table}[t]
\caption{Forecasting Loss of the Hedge Algorithm and HECA}
\label{Table5_new}
\centering
\begin{tabular}{rrrr}
\toprule
& \multicolumn{1}{c}{\citeauthor{FreundSchapire1997}'s} & \multicolumn{1}{c}{HECA} & \multicolumn{1}{c}{Difference}\\
& \multicolumn{1}{c}{Hedge Algorithm} & \multicolumn{1}{c}{Algorithm~\ref{HECA}} & \\
\cmidrule(r){1-4}
2016Q4                   & 0.0051        & 0.0003        & 0.0048    \\
\textcolor{blue}{2017Q1} & 0.1630        & 0.1000        & 0.0630    \\
2017Q2                   & 0.6184        & 0.7911        & -0.1728   \\
2017Q3                   & 0.8484        & 0.9681        & -0.1197   \\
2017Q4                   & 1.0461        & 1.0032        & 0.0428    \\
\textcolor{blue}{2018Q1} & 0.4303        & 0.3646        & 0.0657    \\
2018Q2                   & 0.0372        & 0.0476        & -0.0103   \\
2018Q3                   & 0.2904        & 0.3592        & -0.0688   \\
2018Q4                   & 0.9266        & 0.9614        & -0.0349   \\
\textcolor{blue}{2019Q1} & 0.5574        & 0.6747        & -0.1174   \\
2019Q2                   & 0.5640        & 0.4997        & 0.0643    \\
2019Q3                   & 0.1269        & 0.1667        & -0.0398   \\
2019Q4                   & 0.1434        & 0.0827        & 0.0607    \\
\textcolor{blue}{2020Q1} & 19.3987       & 19.3347       & 0.0640    \\
2020Q2                   & 250.1459      & 248.8660      & 1.2799    \\
2020Q3                   & 29.0959       & 28.7480       & 0.3479    \\[0.6cm]
& \multicolumn{1}{c}{\citeauthor{FreundSchapire1997}'s} & \multicolumn{1}{c}{HECA} & \multicolumn{1}{c}{Difference}\\
& \multicolumn{1}{c}{Hedge Algorithm} & \multicolumn{1}{c}{Algorithm~\ref{HECA_delayed_announcements}} & \\
\cmidrule(r){1-4}
2016Q2                   & 0.0060        & 0.0003        & 0.0057    \\
2016Q3                   & 0.0056        & 0.0053        & 0.0003    \\
2016Q4                   & 0.0053        & 0.0023        & 0.0030    \\
\textcolor{blue}{2017Q1} & 0.1621        & 0.2629        & -0.1008   \\
2017Q2                   & 0.6176        & 0.5335        & 0.0841    \\
2017Q3                   & 0.8499        & 0.7981        & 0.0518    \\
2017Q4                   & 1.0464        & 0.9484        & 0.0980    \\
\textcolor{blue}{2018Q1} & 0.4281        & 0.2621        & 0.1661    \\
2018Q2                   & 0.0389        & 0.0926        & -0.0537   \\
2018Q3                   & 0.2837        & 0.4149        & -0.1312   \\
2018Q4                   & 0.9227        & 0.9729        & -0.0502   \\
\textcolor{blue}{2019Q1} & 0.5284        & 0.6253        & -0.0970   \\
2019Q2                   & 0.5674        & 0.4996        & 0.0677    \\
2019Q3                   & 0.1248        & 0.1241        & 0.0007    \\
2019Q4                   & 0.1419        & 0.0905        & 0.0514    \\
\textcolor{blue}{2020Q1} & 19.3722       & 19.3059       & 0.0663    \\
2020Q2                   & 249.5742      & 244.8226      & 4.7516    \\
2020Q3                   & 29.0456       & 27.9170       & 1.1286
\end{tabular}
\end{table}